\documentclass[aps,amsmath, amssymb, showpacs, showkeys, longbibliography]{revtex4-2}
\usepackage[dvips]{graphicx}
\usepackage{times}
\usepackage{braket}
\usepackage{mathrsfs}
\usepackage{xcolor}
\usepackage{orcidlink}
\usepackage{hyperref}
\hypersetup{
	colorlinks=true,
	urlcolor=blue,
	linkcolor=red,
	citecolor=blue
}
\usepackage{float}
\usepackage{multirow}
\usepackage{enumitem}
\usepackage{multirow}
\usepackage{booktabs}
\usepackage[sort&compress]{natbib}
\usepackage[shortcuts]{extdash}

\begin{document}
\title{Five-Dimensional Traversable Wormholes in Einstein-Gauss-Bonnet Gravity 
with a Cloud of Strings}

\author{Jaydeep Goswami\orcidlink{0009-0007-9467-5664}}
\email[Email: ]{jdpgsm98@gmail.com}

\author{Umananda Dev Goswami\orcidlink{0000-0003-0012-7549}}
\email[Email: ]{umananda@dibru.ac.in}

\affiliation{Department of Physics, Dibrugarh University, Dibrugarh 786004, 
Assam, India}


\begin{abstract}
We present an exact static and spherically symmetric traversable wormhole 
solution in 5D Einstein-Gauss-Bonnet (EGB) gravity supported by a Letelier 
cloud of strings. The physically admissible parameter space is determined by 
imposing the flare-out 
condition together with the positivity of the string-cloud density, and the 
resulting spacetime is shown to be asymptotically flat. An analysis of the 
Ricci scalar, Ricci tensor squared, and Kretschmann scalar confirms that the 
geometry is free from curvature singularities. The effective null, weak, 
strong and dominant energy conditions are either satisfied or saturated at the 
throat. The generalized TOV equation demonstrates that the wormhole remains 
in mechanical equilibrium. We further investigate its traversability by 
studying the proper radial distance, embedding diagrams, traversal time, 
proper acceleration, and tidal accelerations, showing that the solution 
satisfies the Morris-Thorne criteria for traversable wormholes. The optical 
properties of the spacetime are explored through null geodesics, unstable 
photon circular orbits, and the corresponding shadow, revealing the influence 
of both the Gauss-Bonnet coupling and the string-cloud density. Finally, 
scalar perturbations are analyzed using the sixth-order WKB approximation 
method together with time-domain evolution. The quasinormal mode spectra 
exhibit negative imaginary frequencies throughout the considered parameter 
space, indicating linear stability, while the close agreement between the WKB, 
time-domain, and eikonal results provides additional consistency for the 
analysis. These results demonstrate that higher-curvature effects in 5D EGB 
gravity can support a regular, traversable and dynamically stable wormhole 
sustained by a physically motivated string-cloud matter source.
\end{abstract}



\maketitle
\section{Introduction}\label{sec1}
Wormholes are among the most fascinating predictions of gravitational physics, 
representing hypothetical tunnels connecting distant regions of spacetime or 
even separate Universes through a non-trivial topology. The concept was first 
introduced by Wheeler in the context of spacetime quantum foam in 
1957~\cite{a1}, while the earliest mathematical realization appeared in the 
Einstein–Rosen bridge proposed by Einstein and Rosen in 1935~\cite{a2}. 
Although these early solutions were not traversable, the pioneering work of 
Morris and Thorne in 1988~\cite{a3} demonstrated that traversable wormholes 
could exist provided certain geometrical conditions are satisfied. These 
include the absence of event horizons, the flare-out condition at the throat 
and asymptotic flatness, ensuring safe passage through the 
wormhole~\cite{a3,a4,a5,a6,a7,a8,a9,a10}. Owing to these remarkable 
properties, traversable wormholes have attracted sustained interest in 
gravitational physics, astrophysics and cosmology, where they have been 
investigated as possible shortcuts through spacetime, black hole mimickers 
and potential sources of observable signatures~\cite{a11,a12,a13,a14,a15}.

Despite their appealing properties, traversable wormholes within General 
Relativity (GR) suffer from a fundamental difficulty. The flare-out condition 
inevitably requires the violation of the null energy condition, implying that 
the throat must be supported by exotic matter~\cite{a3,a4}. Since no 
observational evidence for such matter currently exists, identifying 
physically realistic mechanisms capable of sustaining traversable wormholes 
remains one of the central challenges in wormhole physics. Various 
possibilities have therefore been explored, including quantum effects, 
non-standard matter sources and extensions of Einstein's theory of 
gravity, i.e., GR~\cite{a16,a17,a18}. Modified theories of gravity (MTGs) that 
arise from the geometric modification of GR, and alternative theories of
gravity (ATGs) that have different geometric structures from GR,  provide a 
natural framework for overcoming this difficulty, because the additional 
geometric contributions appear in the gravitational field equations of such 
theories can effectively sustain the wormhole geometry with only the ordinary 
matter threading the throat to satisfy, or partially satisfy, the classical 
energy conditions. Consequently, wormhole solutions have been extensively 
investigated in several MTGs and ATGs, including Rastall gravity~\citep{a9}, 
$f(R)$ gravity~\cite{a19,a20,a21}, Gauss-Bonnet gravity~\citep{a22} and 
teleparallel gravity~\citep{a23,a24,a25}. These studies demonstrate that 
modifications to the gravitational sector can significantly alter the matter 
requirements needed to maintain traversable wormholes.

Among higher-curvature theories, Einstein–Gauss–Bonnet (EGB) gravity occupies 
a distinguished position as it represents the second-order member of Lovelock 
gravity~\cite{a26} and naturally arises as the leading higher-curvature 
correction in the low-energy limit of heterotic string theory~\cite{a27,a28}. 
Unlike in four dimensions, where the Gauss–Bonnet invariant contributes only 
as a topological term, it becomes dynamically significant in 
higher-dimensional spacetimes~\cite{a29,a30}, leading to a rich variety of 
black hole, cosmological and wormhole solutions~\cite{a27,a31}. Besides the 
gravitational theory, the choice of matter source also plays an essential role 
in determining the physical properties of wormhole spacetimes. One 
particularly interesting matter distribution is the cloud of strings proposed 
by Letelier, which describes a continuous distribution of one-dimensional 
strings~\cite{a32}. Since its introduction, this model has been successfully 
employed in studies of black holes, compact objects and modified gravity, 
where it has been shown to produce significant modifications to the spacetime 
geometry~\cite{a33,a34,a35}.

Besides constructing geometrically consistent wormhole solutions, it is 
equally important to investigate their observational signatures. The motion 
of photons around the wormhole geometry determines the photon-sphere, critical 
impact parameter and shadow radius, providing direct information about the 
underlying spacetime geometry~\cite{a36,a37,a38,a38a}. Recent advances in 
very-long-baseline interferometry~\cite{a39,a40} and gravitational wave 
astronomy~\cite{a41} have made these optical characteristics increasingly 
relevant for testing strong-field gravity and distinguishing wormholes from 
black holes.

The response of a wormhole to external perturbations is encoded in its 
quasinormal modes, which describe the characteristic damped oscillations of 
its spacetime~\cite{a42,a43}. These oscillations depend only on the background 
geometry and therefore provide an important tool for probing compact objects 
and testing gravitational theories~\cite{a43,a44}. Previous investigations 
have shown that wormhole quasinormal modes, echoes and related phenomena may 
provide observational signatures capable of distinguishing wormholes from 
black holes~\cite{a12,a45,a45a}. Consequently, the study 
of quasinormal spectra has become an indispensable component of modern 
wormhole research.

Motivated by these developments, we investigate an exact static and 
spherically symmetric traversable wormhole solution in 5D EGB gravity 
supported by a cloud of strings. We first identify the physically viable 
parameter space through the flare-out condition and the requirement of a 
positive string-cloud density. The resulting spacetime is shown to be 
asymptotically flat and free from curvature singularities by examining the 
Ricci scalar, Ricci tensor squared, and Kretschmann scalar. We then recast 
the field equations of the theory into the form of the effective Einstein 
field equations to investigate the effective energy conditions and the 
mechanical equilibrium of the wormhole, highlighting the role of the 
Gauss–Bonnet curvature terms in sustaining the geometry while the string-cloud 
matter remains physically reasonable. To assess the physical viability of the 
solution, we examine its traversability through the proper radial distance, 
embedding diagrams, traversal time, proper acceleration and radial and lateral 
tidal accelerations. We also explore the optical properties of the wormhole by 
analyzing null geodesics, unstable photon orbits, the photon-sphere and the 
corresponding shadow. Finally, we study the response of the spacetime to 
scalar perturbations using the Pad\'{e}-averaged sixth-order WKB approximation 
together with time-domain evolution, and examine the eikonal correspondence 
between quasinormal modes and unstable photon orbits. Taken together, these 
analyses provide a comprehensive assessment of the geometric, physical, 
observational and dynamical properties of the proposed wormhole solution.

The remainder of this paper is organized as follows. In 
Sec.~\ref{sec2}, we construct the exact 5D EGB wormhole solution supported 
by a cloud of strings and identify the corresponding physically admissible 
parameter space. Sec.~\ref{sec3} examines the regularity of spacetime 
through an analysis of the relevant curvature invariants. In Sec.~\ref{sec4}, 
we investigate the effective energy conditions and the mechanical equilibrium 
of the wormhole. The traversability of the solution is discussed in 
Sec.~\ref{sec5}, followed by an analysis of its optical properties in 
Sec.~\ref{sec6}, including null geodesics, the photon sphere and the shadow. 
Sec.~\ref{sec7} is devoted to quasinormal mode analysis. We first study 
the eikonal quasinormal modes and their relation to unstable photon orbits in 
Sec.~\ref{sec7sub1}, followed by an analysis of scalar quasinormal modes in 
Sec.~\ref{sec7sub2}. The time-domain evolution of scalar perturbations and 
the extraction of the associated quasinormal frequencies are presented in 
Sec.~\ref{sec7sub3}. Finally, Sec.~\ref{sec8} summarizes the main results and 
conclusions of this work.

\section{Five-Dimensional Einstein-Gauss-Bonnet Wormhole Solution}\label{sec2}
We begin with the action for the 5-dimensional (5D) EGB gravity theory with 
matter field in the absence of cosmological constant $(\Lambda=0)$ as given 
by~\cite{a46}
\begin{equation}
S=\int d^5 x \sqrt{-g}\left[\frac{1}{2 \kappa_5^2} (R+\alpha L_\text{GB})\right]+S_\text{matter},
\label{eq1}
\end{equation}
where $\kappa_5=\sqrt{8 \pi G_5}$ with $G_5$ denoting the 5D gravitational 
constant, $R$ is the 5D Ricci scalar and $S_\text{matter}$ is the action for 
matter fields. The second order Gauss-Bonnet term $L_\text{GB}$ is given by
\begin{equation}
L_\text{GB} = R^2-4 R_{\mu\nu}R^{\mu\nu}+R_{\mu\nu\rho\sigma}R^{\mu\nu\rho\sigma}.
\label{eq2}
\end{equation}
The parammeter $\alpha$ is a coupling constant and we assume $\alpha > 0$ 
throughout the study. For $\alpha \to 0$, the field equations reduce to those 
of 5D GR. $R_{\mu\nu}$ and $R_{\mu\nu\rho\sigma}$ are the Ricci tensor and 
Riemann tensor, respectively. Varying the action \eqref{eq1} with respect to 
the metric $g_{\mu\nu}$ and considering $\kappa_5=1$ yields the EGB field 
equations as
\begin{equation}
G_{\mu\nu}+\alpha H_{\mu\nu}= T_{\mu\nu},
\label{eq3}
\end{equation}
where $G_{\mu\nu}$ is the Einstein tensor and $H_{\mu\nu}$ is the Lovelock 
tensor, which explicitly can be expressed as
\begin{equation}
H_{\mu\nu} = 2\left(R R_{\mu\nu}- 2 R_{\mu\alpha} R^\alpha_\nu - 2R^{\alpha\beta} R_{\mu\alpha\nu\beta} + R^{\alpha\beta\gamma}_{\mu} R_{\nu\alpha\beta\gamma}\right) - \frac{1}{2}\, g_{\mu\nu} L_\text{GB}
\label{eq4}
\end{equation}
and $T_{\mu\nu}$ is the energy-momentum tensor for matter fields obtained from 
matter action $S_\text{matter}$. The matter source considered in this work 
corresponds to a Letelier-type cloud of strings~\cite{a32}. In a static and 
spherically symmetric 5D spacetime, the corresponding energy-momentum tensor 
can be written as~\cite{a34}
\begin{equation}
T^\mu_\nu=\text{diag}\left[-\rho_s(r),-\rho_s(r),0,0,0\right],
\label{eq5}
\end{equation}
where $\rho_s(r)$ denotes the string cloud density. To construct a static and 
spherically symmetric spacetime, we consider the metric ansatz
\begin{equation}
ds^2=-f(r)\, dt^2+ \frac{dr^2}{g(r)}+ r^2 d\Omega_3^2.
\label{eq6}
\end{equation}
Here, $d\Omega_3^2=d\chi^2+\sin^2\chi(d\theta^2+\sin^2\theta\,d\phi^2)$
denotes the line element of a unit three-sphere ($S^3$). The conservation 
condition, $\nabla_\mu T^\mu_{\ \nu}=0$, for the energy-momentum tensor 
\eqref{eq5} yields the radial equation:
\begin{equation}
\rho_s'(r)+\frac{3}{r}\rho_s(r)=0.
\label{eq7}
\end{equation}
Integrating this Eq.~\eqref{eq7}, one can obtain
\begin{equation}
\rho_s(r)=\frac{\rho_0}{r^3},
\label{eq8}
\end{equation}
where $\rho_0$ is an integration constant characterizing the string cloud 
density. Since a physically reasonable matter distribution requires 
$\rho_s(r)\ge 0$ for $r>0$, it follows that $\rho_0 \geq 0$. The special case 
$\rho_0=0$ corresponds to the absence of the string cloud, reducing the matter 
sector to vacuum. Substituting the metric ansatz~\eqref{eq6} and the 
string-cloud energy-momentum tensor~\eqref{eq5} into the field equations
\eqref{eq3}, we obtain
\begin{align}
6\alpha\big[1-g(r)\big]g'(r)+\frac{3 r}{2}\big[rg'(r)-2\big(1-g(r)\big)\big] & = - \rho_0,\label{eq9}\\[5pt]
3r\big[1-g(r)\big]- \frac{3f'(r)g(r)}{2f(r)}\big[4\alpha\big(1-g(r)\big) + r^2\big] & = \rho_0,\label{eq10}\\[5pt]
\left[\frac{f''(r)g(r)}{2f(r)} - \frac{f'^2(r)g(r)}{4f^2(r)}\right]\big[4\alpha\big(1-g(r)\big) + r^2\big]&\nonumber\\ 
+\, \frac{f'(r)}{f(r)}\left[\frac{g'(r)}{4}\big\{4\alpha\big(1-3g(r)\big) + r^2\big\} + rg(r)\right] + rg'(r) + g(r) -1 & = 0.
\label{eq11}
\end{align}
Here, prime $(')$ denotes the derivative of the function with respect to the 
radial coordinate $r$. Introducing $h(r)=1-g(r)$ and hence $h'(r)=-g'(r)$, 
Eq.~\eqref{eq9} can be written in the form: $6 \alpha (h^2)' + 3 (r^2h)' = 
2 \rho_0$. Integrating this, one may obtain
\begin{equation}
6 \alpha h^2 + 3 r^2 h - 2 \rho_0 r - C_g = 0,
\label{eq12}
\end{equation}
where $C_g$ is an integration constant. Solving Eq.~\eqref{eq12} for $h(r)$ 
yields
\begin{equation}
h(r) = -\,\frac{r^2}{4 \alpha}\left[1 \mp \sqrt{1+ \frac{16 \alpha \rho_0}{3r^3}+\frac{8 \alpha C_g}{3 r^4}}\;\right]\!.
\label{eq13}
\end{equation}
Hence, we get
\begin{equation}
g(r) =
1+ \frac{r^2}{4 \alpha}\left[1 \mp \sqrt{1+ \frac{16 \alpha \rho_0}{3r^3}+\frac{8 \alpha C_g}{3 r^4}}\;\right]\!.
\label{eq14}
\end{equation}
Imposing the throat condition, $g(r_0)=0$~\cite{a3}, yields
\begin{equation}
C_g=6 \alpha +3 r_0^2-2 \rho_0 r_0.
\label{eq15}
\end{equation}
Using this value of $C_g$ in Eq.~\eqref{eq14}, we get
\begin{equation}
g(r) = 1 + \frac{r^2}{4 \alpha}\left[1 \mp \sqrt{1+ \frac{16 \alpha \rho_0}{3r^3}+\frac{8 \alpha}{3 r^4} (6 \alpha +3 r_0^2-2 \rho_0 r_0)}\right].
\label{eq16}
\end{equation}
Again, rearranging Eq.~\eqref{eq9} we can have
\begin{equation}
g'(r)= \frac{2 \big(3r -3rg(r) -\rho_0\big)}{3 \big(r^2 -4 \alpha g(r) + 4 \alpha\big)}.
\label{eq17}
\end{equation}
Evaluating this Eq.~\eqref{eq17} at the throat and imposing the flaring-out 
condition, $g'(r_0)>0$~\cite{a16}, we get
\begin{equation}
\frac{6 r_0 - 2 \rho_0}{3r_0^2 +12 \alpha}>0.
\label{ineq18}
\end{equation}
The denominator in the inequality Eq.~\eqref{ineq18} is positive for 
$\alpha>-\,r_0^2/4$, and also the flaring-out condition requires that 
$\rho_0<3r_0$. Combining these constraints with the assumptions of a 
non-negative Gauss-Bonnet coupling $\alpha$ and a physically acceptable string 
cloud density ($\rho_s>0$), the physically admissible parameter space is 
constrained to $\alpha > 0$ and $0 \leq \rho_0 < 3r_0$. This guarantees both 
a positive string density and satisfaction of the flare-out condition in 5D 
EGB gravity. To recover the corresponding 5D Einstein GR limit solution, we 
examine $\alpha \to 0$. In this limit, the positive branch of $g(r)$ diverges, 
whereas its negative branch remains finite. Therefore, only the negative 
branch admits a smooth GR limit and will be considered henceforth. 
Consequently, Eq.~\eqref{eq16} reduces to
\begin{equation}
g(r) = 1 + \frac{r^2}{4 \alpha}\left[1 - \sqrt{1+ \frac{16 \alpha \rho_0}{3r^3}+\frac{8 \alpha}{3 r^4} (6 \alpha +3 r_0^2-2 \rho_0 r_0)}\;\right]\!.
\label{eq19}
\end{equation}
Further, rearranging Eq.~\eqref{eq10} one can wirte
\begin{equation}
\frac{f'(r)}{f(r)}
= \frac{2 \big(3r -3rg(r) -\rho_0\big)}{3\,g(r) \big(r^2 -4 \alpha g(r) + 4 \alpha\big)}
= \frac{g'(r)}{g(r)}.
\label{eq20}
\end{equation}
Integrating Eq.~\eqref{eq20}, one can find $f(r)=f_0 g(r)$, where 
$f_0$ is an integration constant. Now, imposing the asymptotic normalization 
condition, which fixes the normalization of the timelike Killing vector at 
spatial infinity, i.e., $\left.f(r)/g(r)\right|_{r \to \infty}=1$, 
it can be found that $f_0=1$, and therefore $f(r) = g(r)$. Accordingly, the 
metric element in Eq.~\eqref{eq6} becomes
\begin{equation}
ds^2=-f(r)\, dt^2+ \frac{dr^2}{f(r)}+ r^2 \big[d\chi^2+\sin^2\chi(d\theta^2+\sin^2\theta\,d\phi^2)\big],
\label{eq21}
\end{equation}
where
\begin{equation}
f(r)= 1+ \frac{r^2}{4 \alpha}\left[1 - \sqrt{1+ \frac{16 \alpha \rho_0}{3r^3}+\frac{8 \alpha}{3 r^4} (6 \alpha +3 r_0^2-2 \rho_0 r_0)}\;\right].
\label{eq22}
\end{equation}
The spacetime geometry is therefore fully specified by the metric function 
$f(r)$ as given in Eq.~\eqref{eq22} together with the throat radius $r_0$, the 
Gauss–Bonnet coupling $\alpha$ and the string-cloud parameter $\rho_0$. 
Furthermore, for sufficiently large radial distances, the metric function 
$f(r)$ can be approximated as
\begin{equation}
f(r) = 1 - \frac{2 \rho_0}{3 r} - \frac{6 \alpha +3 r_0^2-2 \rho_0 r_0}{3 r^2} + O(r^{-4}),
\label{eq23}
\end{equation}
which implies that in the limit $r \to \infty$, $f(r)=1$. This shows that 
the spacetime approaches the Minkowski limit and is therefore asymptotically 
flat.

It is important to emphasize that the present solution does not admit a direct 
4D-limit within the framework of standard EGB gravity. In four spacetime 
dimensions, the Gauss-Bonnet invariant becomes a topological quantity whose 
variation does not contribute to the gravitational field equations. 
Consequently, the standard EGB theory reduces to GR, and the higher-curvature 
corrections responsible for the present 5D solution disappear. To circumvent 
this difficulty, Glavan and Lin proposed a regularization procedure~\cite{a47} 
in which the Gauss-Bonnet coupling is rescaled as
$$\alpha \rightarrow \frac{\alpha}{D-4},$$
in the limit $D\rightarrow4$. This prescription yields an effective 
4D EGB theory with nontrivial contributions from the Gauss-Bonnet term. 
However, the resulting theory is conceptually distinct from the genuine 5D EGB 
framework considered here and has been the subject of considerable discussion 
in the literature. Therefore, the present wormhole solution should be regarded 
as an intrinsically 5D configuration and does not possess a direct 4D 
counterpart obtained through conventional dimensional reduction.

\begin{figure}[!h]
    \includegraphics[scale=0.65]{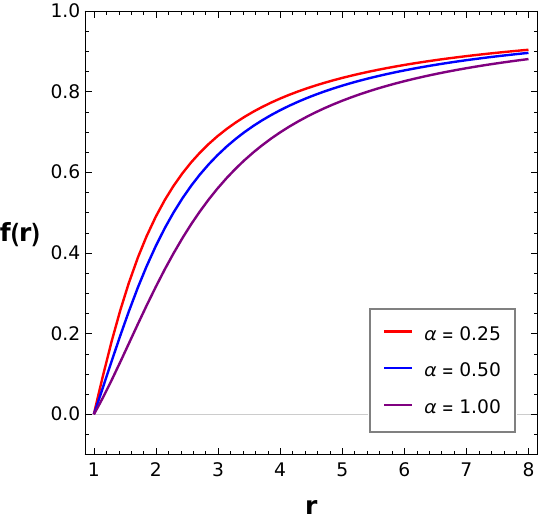}\hspace{1.0cm}
    \includegraphics[scale=0.65]{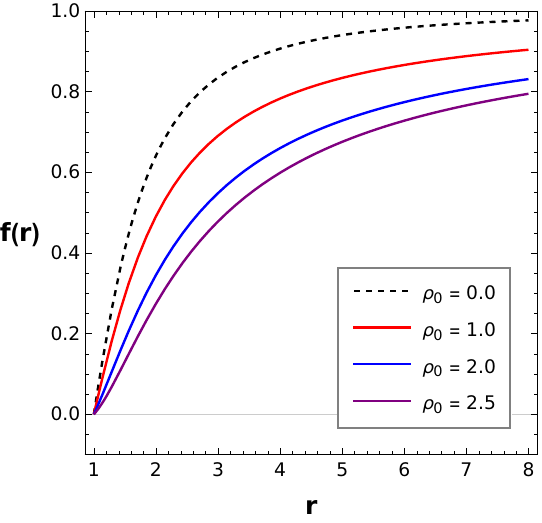}
    \vspace{-0.2cm}
    \caption{Metric function $f(r)$ of the wormhole spacetime for different 
values of the Gauss–Bonnet coupling parameter $\alpha$ with fixed $\rho_0=1$ 
(left panel) and matter density parameter $\rho_0$ with fixed $\alpha=0.25$ 
(right panel) defined by Eq.~\eqref{eq22} with the throat radius $r_0 = 1$.}
    \label{fig1}
\end{figure}

Fig.~\ref{fig1} illustrates the radial behaviour of the metric function $f(r)$ 
for different values of the Gauss-Bonnet coupling parameter $\alpha$ 
(left panel) and the matter density parameter $\rho_0$ (right panel). In all 
cases, the metric function satisfies $f(r_0)=0$ at the wormhole throat 
($r_0=1$) and increases monotonically with the radial coordinate, 
asymptotically approaching $f(r)\rightarrow1$ as $r\rightarrow\infty$. This 
confirms that the obtained wormhole solutions are asymptotically flat. For a
fixed matter density ($\rho_0=1$), the metric function decreases as $\alpha$ 
increases throughout the spacetime and approaches its asymptotic value more 
gradually. The ordering
$f_{\alpha\,=\,0.25}(r) > f_{\alpha\,=\,0.5}(r) > f_{\alpha\,=\,1}(r)$ for 
all $r>r_0$ indicates that the higher-curvature Gauss-Bonnet corrections 
effectively suppress the function, making the geometry more strongly curved 
near the throat while preserving asymptotic flatness. For a fixed Gauss-Bonnet 
coupling ($\alpha=0.25$), increasing the matter density parameter $\rho_0$ 
also lowers the metric function, with the vacuum solution ($\rho_0=0$) 
exhibiting the largest values of $f(r)$. The ordering
$f_{\rho_0\,=\,0}(r) > f_{\rho_0\,=\,1}(r) > f_{\rho_0\,=\,2}(r) > 
f_{\rho_0\,=\,2.5}(r)$
for all $r>r_0$ shows that the presence of the string cloud strengthens the 
gravitational field and produces a more pronounced redshift effect in the 
vicinity of the wormhole throat. Overall, both the Gauss-Bonnet coupling and 
the matter density suppress the metric function near the throat while 
preserving its smooth and monotonic behaviour. Nevertheless, all solutions 
approach $f(r)=1$ at large distances, demonstrating that the combined effects 
of higher-curvature corrections and the string cloud modify only the local 
geometry without spoiling the asymptotic flatness of the spacetime.

\section{Curvature properties and regularity}\label{sec3}

\begin{figure}[!h]
    \includegraphics[scale=0.7]{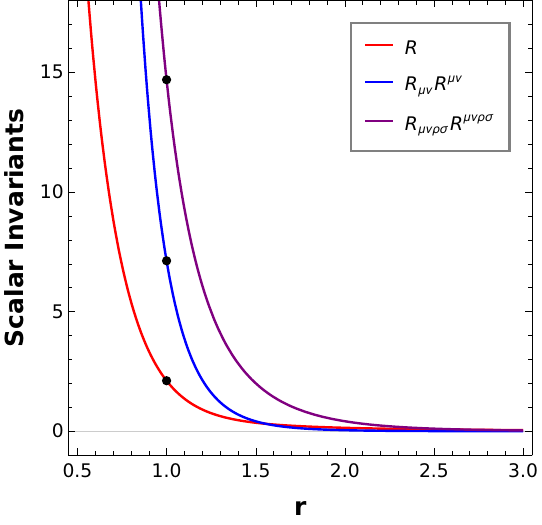}
    \vspace{-0.2cm}
    \caption{Behaviours of Ricci scalar $R$, the Ricci tensor contraction 
$R_{\mu\nu}R^{\mu\nu}$ and the Kretschmann scalar 
$R_{\mu\nu\rho\sigma}R^{\mu\nu\rho\sigma}$ with respect to $r$ for 
Gauss-Bonnet coupling parameter $\alpha=0.25$, matter density parameter 
$\rho_0=1$ and throat radius $r_0=1$. The black dots in the plot correspond to 
the throat radius values.}
    \label{fig2}
\end{figure}

In this section, we examine the spacetime curvature through scalar invariants 
constructed from the Riemann tensor to verify the regularity of the obtained 
wormhole geometry. Since these invariants are independent of the choice of 
coordinates, they provide a reliable criterion for identifying genuine 
curvature singularities~\cite{a48}. In particular, we evaluate the Ricci 
scalar $R$, the Ricci tensor squared $R_{\mu\nu}R^{\mu\nu}$, and the 
Kretschmann scalar $R_{\mu\nu\rho\sigma}R^{\mu\nu\rho\sigma}$. If these 
quantities remain finite throughout the spacetime, especially at the wormhole 
throat, the geometry is free from physical curvature singularities. 
Fig.~\ref{fig2} shows the radial behavior of these curvature invariants for 
the representative parameter values $\alpha=0.25$, $\rho_0=1$ and $r_0=1$. It 
is evident that all invariants remain finite at the throat and decrease 
smoothly with increasing radial distance, approaching zero in the asymptotic 
region. The regular behavior of these curvature scalars confirms that the 
wormhole spacetime is free from curvature singularities. Furthermore, the 
vanishing of the invariants at large $r$ is consistent with the asymptotically 
flat nature of the obtained solution.

\begin{figure}[!h]
    \includegraphics[scale=0.65]{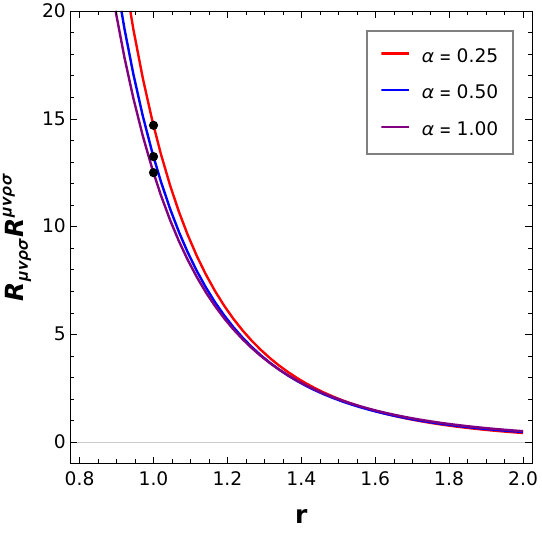}\hspace{1.0cm}
    \includegraphics[scale=0.65]{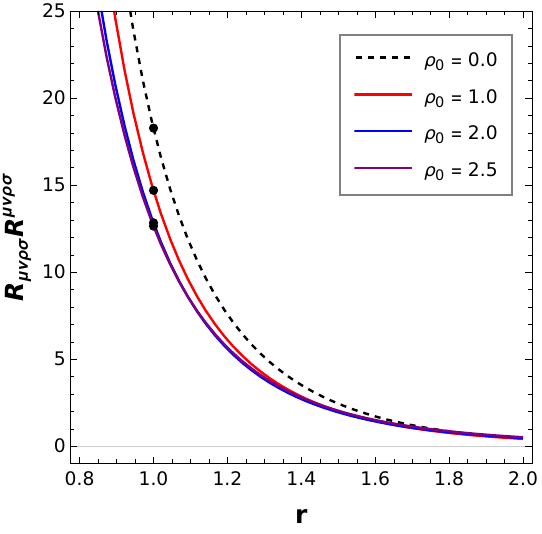}
    \vspace{-0.2cm}
    \caption{Kretschmann scalar $R_{\mu\nu\rho\sigma}R^{\mu\nu\rho\sigma}$ for 
different values of the Gauss–Bonnet coupling parameter $\alpha$ with fixed 
$\rho_0=1$ (left panel) and matter density parameter $\rho_0$ with fixed 
$\alpha=0.25$ (right panel) with the throat radius $r_0 = 1$.The black dots in 
both plots correspond to values at the throat radius.}
    \label{fig3}
\end{figure}

To understand the individual roles of the Gauss-Bonnet coupling and the matter 
distribution, we restrict our parameter analysis to the Kretschmann scalar in 
Fig.~\ref{fig3}, which provides a comprehensive measure of the spacetime 
curvature. In the left panel, for a fixed matter density $\rho_0=1$, 
increasing the Gauss-Bonnet coupling $\alpha$ leads to a slight suppression 
of the Kretschmann scalar in the vicinity of the wormhole throat. However, 
the differences rapidly diminish with increasing radial distance, and all 
curves converge asymptotically to zero. This indicates that the Gauss-Bonnet 
corrections primarily influence the local curvature near the throat while 
leaving the asymptotic structure essentially unchanged. The right panel 
depicts the variation of the Kretschmann scalar with the matter density 
parameter $\rho_0$ for a fixed coupling $\alpha=0.25$. The vacuum solution 
($\rho_0=0$) possesses the largest curvature near the throat, whereas 
increasing $\rho_0$ systematically reduces the magnitude of the Kretschmann 
scalar in the near-throat region. Nevertheless, all curves remain finite 
throughout the physical domain and decay monotonically to zero as 
$r \to \infty$, confirming that the spacetime remains regular and 
asymptotically flat for the considered parameter range.
To further verify the regularity of the wormhole geometry, we evaluate the 
curvature invariants at the throat. For the representative parameter set 
$(r_0,\rho_0,\alpha)=(1,1,0.25)$, the corresponding values of the Ricci 
scalar, Ricci tensor squared and Kretschmann scalar are $R(r_0)=2.1111$, 
$R_{\mu\nu}R^{\mu\nu}(r_0)=7.1173$ and 
$R_{\mu\nu\rho\sigma}R^{\mu\nu\rho\sigma}(r_0)=14.6790$ respectively. The 
finiteness of all these invariants at $r=r_0$ confirms that the wormhole 
throat is a regular hypersurface rather than a curvature singularity.

\section{Effective energy conditions and mechanical equilibrium}\label{sec4}

Energy conditions provide a set of mathematical constraints on the 
energy-momentum tensor that describe the behavior of matter and energy in a
given spacetime. These usual conditions are: null energy condition (NEC),
weak energy condition (WEC), strong energy condition (SEC) and dominant energy 
condition (DEC), and they are closely related to the Raychaudhuri equation, 
which governs the focusing of timelike and null geodesic 
congruences~\cite{a49}. Within the framework of GR, physically reasonable 
matter fields are expected to satisfy these energy conditions~\cite{a3}. For 
the string cloud energy-momentum tensor~\eqref{eq5}, the energy conditions are 
automatically satisfied or saturated provided $\rho_s \geq 0$. Therefore, the 
supporting matter is non-exotic~\cite{a50,a51}. Any effective deviation from 
the classical energy conditions arises solely from the Gauss-Bonnet curvature 
contributions when the field equations are recast in Einstein form. To examine 
the effective energy conditions, we defined the effective energy-momentum 
tensor using Eq.~\eqref{eq3} as 
$T_{\mu\nu}^{\text{eff}} = T_{\mu\nu} - \alpha H_{\mu\nu}$ and consequently, 
the field equations take the Einstein-like form 
$G_{\mu\nu}=T_{\mu\nu}^{\text{eff}}$~\cite{a52}. It is important to emphasize 
that $T_{\mu\nu}^{\text{eff}}$ is an effective quantity obtained by absorbing 
the Gauss–Bonnet contributions into the matter sector. The physical matter 
source remains the string cloud described by Eq.~\eqref{eq5}, which itself 
satisfies the standard energy conditions. Thus, considering a 5D 
anisotropic effective energy-momentum tensor
\begin{equation}
T_{\mu\nu}^{\text{eff}}
= (\rho^{\text{eff}} + p_t^{\text{eff}}) u_\mu u_\nu- p_t^{\text{eff}} 
g_{\mu\nu} + (p_r^{\text{eff}} - p_t^{\text{eff}}) v_\mu v_\nu,
\label{eq24}
\end{equation}
where $u_\mu$ is the five-velocity, $v_\mu$ is a unit spacelike vector in the 
radial direction, $g_{\mu\nu}$ is the metric tensor defined by 
Eq.~\eqref{eq21}, $\rho^{\text{eff}}$, $p_r^{\text{eff}}$ and 
$p_t^{\text{eff}}$ are the effective energy density, radial pressure and 
tangential pressure, respectively, the effective energy conditions are defined 
as follows~\cite{a53,a54}.
The NEC demands that for any null vector $k^\mu$, the effective 
energy-momentum tensor $T_{\mu\nu}^{\,\text{eff}}$ should be such that
$T_{\mu\nu}^{\text{eff}}\,k^\mu k^\nu \geq 0$ with $k^\mu k_\mu = 0.$
For an anisotropic fluid, this condition reduces to
$\rho^{\text{eff}} + p_r^{\text{eff}} \geq 0$ and 
$\rho^{\text{eff}} + p_t^{\text{eff}} \geq 0.$
The NEC is commonly regarded as a minimal requirement for classical matter.
Whereas the WEC requires that for any timelike vector $u^\mu$, 
$T_{\mu\nu}^{\,\text{eff}}$ has to satisfy
$T_{\mu\nu}^{\text{eff}}\,u^\mu u^\nu \geq 0$ with $u^\mu u_\mu = -\,1.$
For an anisotropic fluid, this condition implies that
$\rho^{\text{eff}} \geq 0,$
$\rho^{\text{eff}} + p_r^{\text{eff}} \geq 0$ and
$\rho^{\text{eff}} + p_t^{\text{eff}} \geq 0.$
The WEC ensures that all observers measure a non-negative energy density.
Similarly, according to the SEC, for any timelike vector $u^\mu$,
$\left(T_{\mu\nu}^{\text{eff}} - 
\frac{1}{3}\, g_{\mu\nu} T^{\text{eff}}\right) u^\mu u^\nu \geq 0,$ 
where $T^{\text{eff}} = g^{\mu\nu} T_{\mu\nu}^{\text{eff}}$. For an 
anisotropic fluid, this gives
$2\,\rho^{\text{eff}} + p_r^{\text{eff}} + 3\,p_t^{\text{eff}} \geq 0.$
The SEC is associated with the focusing of timelike geodesics.
Again, the DEC claim that for any timelike vector $u^\mu$,
$T_{\mu\nu}^{\text{eff}}\,u^\mu u^\nu \geq 0$ with
$T_{\mu\nu}^{\text{eff}}\,u^\mu < 0$. For an anisotropic fluid, this leads 
to $\rho^{\text{eff}} - |p_r^{\text{eff}}| \geq 0$ and
$\rho^{\text{eff}} - |p_t^{\text{eff}}|\geq 0.$

Using Eq.~\eqref{eq24} and the Einstein-like field equations, expressions
for $\rho^{\text{eff}}(r)$, $p_r^{\text{eff}}(r)$ and $p_t^{\text{eff}}(r)$ 
can be obtained as
\begin{align}
\label{eq25a}
\rho^{\text{eff}}(r) & = -\frac{3\big[rf'(r)+2f(r)-2\big]}{2r^2},\\
p_r^{\text{eff}}(r) & = \frac{3\big[rf'(r)+2f(r)-2\big]}{2r^2},\\
p_t^{\text{eff}}(r)&=
\frac{f''(r)}{2}+\frac{2 f'(r)}{r}+\frac{f(r)}{r^2}-\frac{1}{r^2}.
\label{eq25}
\end{align}
Substituting the metric function $f(r)$ from Eq.~\eqref{eq22} in 
Eqs.~\eqref{eq25a}--\eqref{eq25} and evaluating them at the throat $(r=r_0)$, 
we obtain
\begin{align}
\rho^{\text{eff}}(r_0) & =
\frac{12 \alpha + r_0 \rho_0}{r_0^4 + 4 r_0^2 \alpha},
\label{eq26}\\[5pt]
p_r^{\text{eff}}(r_0) & =
-\frac{12 \alpha + r_0 \rho_0}{r_0^4 + 4 r_0^2 \alpha},
\label{eq27}\\[5pt]
p_t^{\text{eff}}(r_0) & =
\frac{4 \alpha \left[ 45 r_0^4 -144 \alpha^2 -24 r_0^3 \rho_0 -48 r_0 \alpha \rho_0 + 2 r_0^2 \left(36 \alpha + \rho_0^2\right)\right]}
{9 r_0 \left(r_0^2 + 4 \alpha \right)^3}.
\label{eq28}
\end{align}
Using these Eqs.~\eqref{eq26}--\eqref{eq28}, the specific expressions for the 
four energy conditions can be obtained as
\begin{align}
& \rho^{\text{eff}}(r_0)+p_r^{\text{eff}}(r_0) = 0,
\label{eq29}\\[5pt]
& \rho^{\text{eff}}(r_0)+p_t^{\text{eff}}(r_0) =
\frac{288\alpha \left(r_0^2+2\alpha\right)^2+3r_0\left(r_0^2-4\alpha\right)\left(3r_0^2+4\alpha\right)\rho_0+8r_0^2\alpha \rho_0^2}
{9r_0^2\left(r_0^2+4\alpha\right)^3},
\label{eq30}\\[5pt]
&2\,\rho^{\text{eff}}(r_0)+p_r^{\text{eff}}(r_0)+3\,p_t^{\text{eff}}(r_0) =
\frac{72 \alpha r_0 \left(3r_0^2+8\alpha\right) +
3\left(r_0^4-24\alpha r_0^2-48\alpha^2\right)\rho_0 +8\alpha r_0 \rho_0^2}{
3r_0\left(r_0^2+4\alpha\right)^3},
\label{eq31}\\[5pt]
&\rho^{\text{eff}}(r_0)-\left|p_r^{\text{eff}}(r_0)\right| = 0,
\label{eq32}\\[5pt]
&\rho^{\text{eff}}(r_0)-\left|p_t^{\text{eff}}(r_0)\right| =
\frac{\left(108\alpha+9r_0\rho_0\right)}{9r_0^2\left(r_0^2+4\alpha\right)}-
\frac{4\alpha\left|45r_0^4-144\alpha^2-24r_0^3\rho_0-48r_0\alpha\rho_0+2r_0^2\left(36\alpha+\rho_0^2\right)\right|}
{9r_0^2\left(r_0^2+4\alpha\right)^3}.
\label{eq33}
\end{align}

\begin{figure}[!h]
    \includegraphics[scale=0.6]{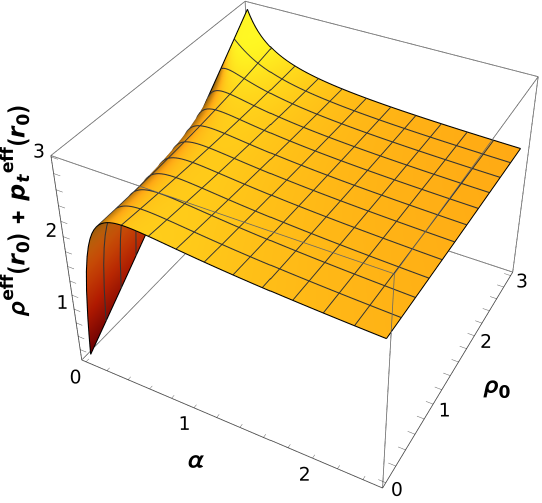}\hspace{0.5cm}
    \includegraphics[scale=0.6]{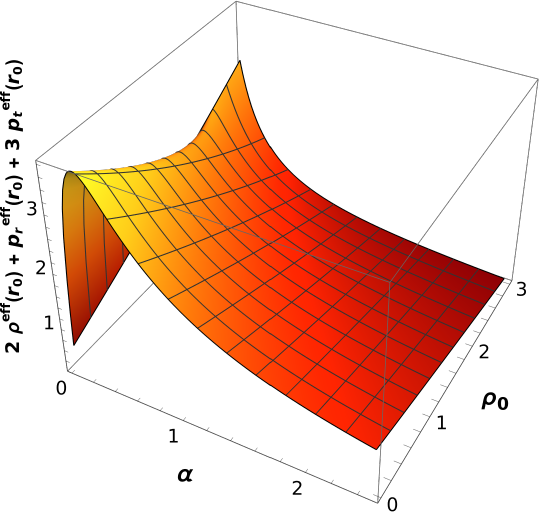}\hspace{0.5cm}
    \includegraphics[scale=0.6]{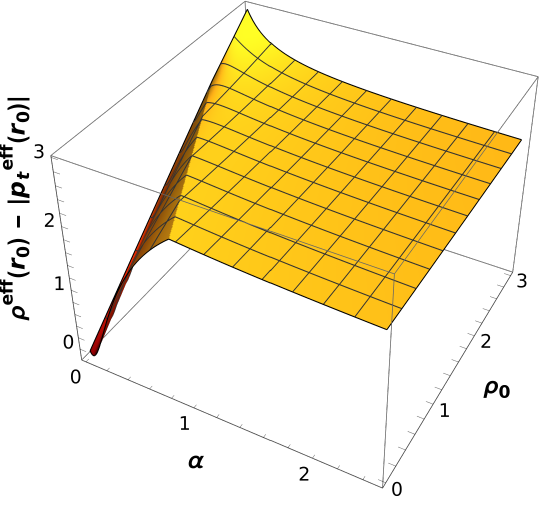}
    \vspace{-0.0cm}
    \caption{3D surfaces illustrating the effective tangential NEC 
$\rho^{\text{eff}}+p_t^{\text{eff}}$ (left panel), the effective SEC 
$2\,\rho^{\text{eff}}+p_r^{\text{eff}}+3\,p_t^{\text{eff}}$ (middle panel) and 
the effective tangential DEC $\rho^{\text{eff}}-|p_t^{\text{eff}}|$ 
(right panel) as functions of the Gauss-Bonnet coupling $\alpha$ and the 
string cloud parameter $\rho_0$ for $r_0=1$.}
\label{fig4}
\end{figure}

From Eq.~\eqref{eq26}, we see that $\rho^{\text{eff}}(r_0)>0$ for $\alpha > 0$ 
and $0 \leq \rho_0 < 3r_0$ at the throat $r=r_0$, implying that the effective 
energy density remains positive at the throat and from Eq.~\eqref{eq29}, the 
radial NEC is identically saturated at the throat. The tangential component of 
the NEC is governed by Eq.~\eqref{eq30}. The corresponding 3D surface, shown 
in Fig.~\ref{fig4} (left panel), remains strictly positive throughout the 
admissible parameter space, demonstrating that $\rho^{\text{eff}}(r_0)+
p_t^{\text{eff}}(r_0)>0$. Consequently, the effective NEC is satisfied at the 
wormhole throat, with the radial component being saturated and the tangential 
component remaining positive. Since $\rho^{\text{eff}}(r_0)>0$, the WEC is 
also satisfied.
The SEC is determined by Eq.~\eqref{eq31}. The middle panel of Fig.~\ref{fig4} 
illustrates the variation of the quantity 
$2\rho^{\text{eff}}+p_r^{\text{eff}}+3p_t^{\text{eff}}$ as a function of the 
Gauss-Bonnet coupling $\alpha$ and the string cloud parameter $\rho_0$ for 
$r_0=1$. The effective SEC remains positive throughout the parameter range 
$\alpha > 0$ and $0 \leq \rho_0 < 3r_0$ at the throat $r=r_0$, indicating that 
this energy condition is satisfied. The magnitude of the effective SEC 
decreases smoothly with increasing $\alpha$ and $\rho_0$, with the largest 
values occurring for smaller values of both parameters. Although the combined 
effect of the Gauss-Bonnet coupling and the string cloud weakens the effective 
SEC, no violation is observed within the explored range of parameters.
The DEC is examined through Eqs.~\eqref{eq32} and \eqref{eq33}. The radial 
component of the DEC is identically saturated, while the tangential component 
remains positive throughout the admissible parameter space, as illustrated in 
Fig.~\ref{fig4} (right panel). Hence, the effective DEC is satisfied at the 
throat of the wormhole.

To examine the mechanical stability of the proposed wormhole, we employ the 
TOV equation for a 5D spacetime, which is given as \cite{a55,a56}
\begin{equation}
-\frac{dp_r^{\text{eff}}}{dr}-\frac{f'(r)}{2 f(r)}\,(\rho^{\text{eff}}+p_r^{\text{eff}})+\frac{3}{r}\,(p_t^{\text{eff}}-p_r^{\text{eff}})=0.
\label{eq37}
\end{equation}
This equation can be decomposed into three distinct forces: the hydrostatic 
force $F_H(r)=-\,dp_r^{\text{eff}}/dr$, gravitational force 
$F_G=-(f'(r)/2 f(r))\left[\rho^{\text{eff}}+p_r^{\text{eff}}\right]$ and 
anisotropic force $F_A=3(p_t^{\text{eff}}-p_r^{\text{eff}})/r$. The wormhole 
is in mechanical equilibrium at the throat when these forces balance each 
other, i.e., $F_H+F_G+F_A=0$. We evaluate the hydrostatic, gravitational and 
anisotropic forces for the wormhole metric~\eqref{eq22} at the throat ($r=r_0$) 
as
\begin{align}
F_H(r_0) & = - \frac{288\alpha\left(r_0^2+2\alpha\right)^2 + 3r_0\left(r_0^2-4\alpha\right)\left(3r_0^2+4\alpha\right)\rho_0 +8r_0^2\alpha\rho_0^2}
{3r_0^3\left(r_0^2+4\alpha\right)^3},
\label{eq38}\\[5pt]
F_G(r_0) & = 0,
\label{eq39}\\[5pt]
F_A(r_0) &  =\frac{288\alpha\left(r_0^2+2\alpha\right)^2 + 3r_0\left(r_0^2-4\alpha\right)\left(3r_0^2+4\alpha\right)\rho_0 +8r_0^2\alpha\rho_0^2}
{3r_0^3\left(r_0^2+4\alpha\right)^3}.
\label{eq40}
\end{align}
Since $\rho^{\text{eff}}(r_0)+p_r^{\text{eff}}(r_0) = 0$ as seen in 
Eq.~\eqref{eq29}, the gravitational force vanishes identically at the throat, 
which leads to Eq.~\eqref{eq39}. From Eqs.~\eqref{eq38} and~\eqref{eq40}, it 
is clear that $F_H(r_0)=-\,F_A(r_0)$ at the throat. Consequently, the 
equilibrium condition $F_H(r_0)+F_G(r_0)+F_A(r_0)=0$ is satisfied at the 
throat without any contribution from the gravitational force.

\section{Traversability analysis}\label{sec5}

In this section, we investigate the physical traversability of the obtained 
wormhole solution. Following the traversability criteria proposed by Morris 
and Thorne~\cite{a3,a4}, we examine whether the wormhole possesses a finite 
proper radial distance, admits a smooth geometric embedding, allows traversal 
within a finite proper time and ensures that the proper acceleration and tidal 
forces experienced by a traveler remain within physically acceptable limits. 
In the following subsections, each of these requirements is analyzed to 
determine whether the obtained 5D EGB wormhole is suitable for safe 
traversal.

\subsection{Proper Radial Distance and Geometric Embedding}\label{sec5sub1}
The proper radial distance is a coordinate-independent quantity that measures 
the spatial distance from the wormhole throat and plays an important role in 
assessing the traversability of the geometry~\cite{a3}. For a given metric 
function $f(r)$, it is expressed as
\begin{equation}
\ell(r)=\pm\int_{r_0}^{r}\frac{dr}{\sqrt{f(r)}},
\label{eq41}
\end{equation}
where the positive and negative branches correspond to the two asymptotically 
flat regions connected by the wormhole. Since our metric function \eqref{eq22} 
does not admit a simple analytical form, the above integral is evaluated 
numerically.

\begin{figure}[!h]
\centering
\includegraphics[scale=0.65]{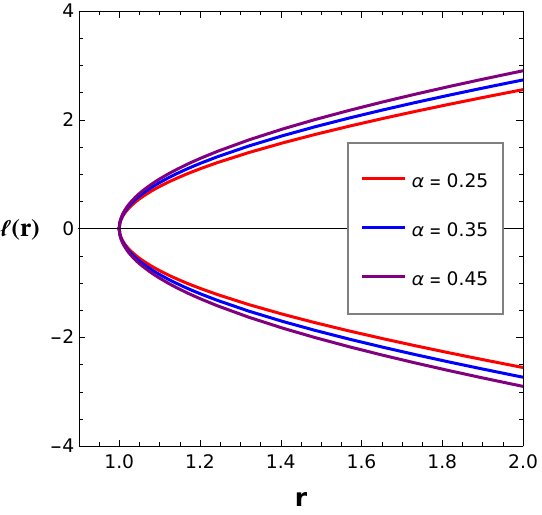}\hspace{1.0cm}
\includegraphics[scale=0.66]{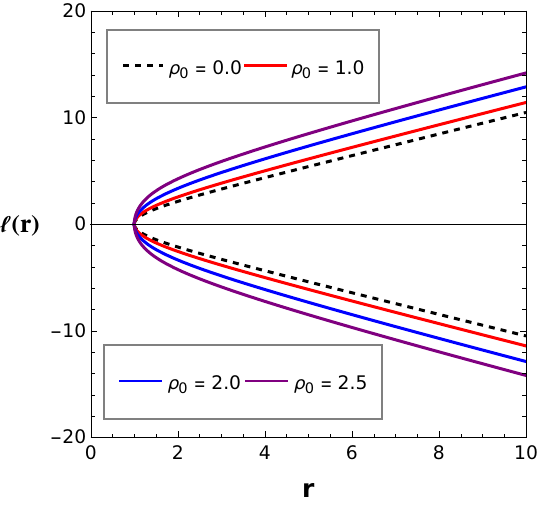}
\vspace{-0.2cm}
\caption{Proper radial distance $\ell(r)$ as a function of the radial coordinate 
$r$ for different values of the Gauss-Bonnet coupling parameter $\alpha$ with 
fixed string cloud density $\rho_0=1$ (left panel) and for different values of 
the string cloud density $\rho_0$ with fixed Gauss-Bonnet coupling 
$\alpha=0.25$ (right panel). In both panels, the throat radius is fixed at 
$r_0=1$.}
\label{fig5}
\end{figure}

Fig.~\ref{fig5} shows the proper radial distance for different values of the 
Gauss-Bonnet coupling parameter $\alpha$ and the string cloud parameter 
$\rho_0$. In all cases, the proper distance vanishes smoothly at the throat, 
$\ell(r_0)=0$, and increases monotonically with the radial coordinate, 
approaching $\pm\infty$ as $r\rightarrow\infty$. This behavior indicates that 
the two asymptotically flat regions are smoothly joined through a regular 
throat without any geometrical discontinuity or singularity. The left panel 
demonstrates that for a fixed radial coordinate, the proper radial distance 
increases with the Gauss-Bonnet coupling $\alpha$, suggesting that 
higher-curvature corrections effectively enlarge the radial extent of the 
wormhole geometry. Similarly, the right panel shows that introducing a cloud 
of strings further increases the proper distance compared with the vacuum case 
($\rho_0=0$). As the string cloud density $\rho_0$ increases, the spatial 
separation measured along the radial direction becomes larger, implying that 
the string cloud enhances the effective size of the wormhole. The smooth and 
well-behaved profile of the proper radial distance throughout the spacetime 
confirms the regularity of the geometry and provides further evidence for the 
traversable nature of the obtained wormhole solution.

\begin{figure}[!h]
\centering
\includegraphics[scale=0.55]{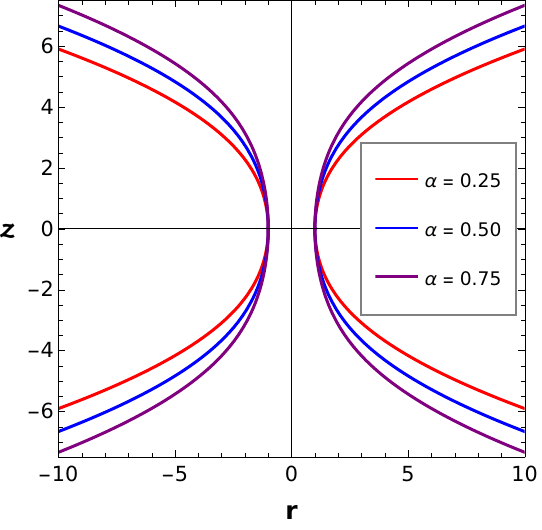}\hspace{0.5cm}
\includegraphics[scale=0.55]{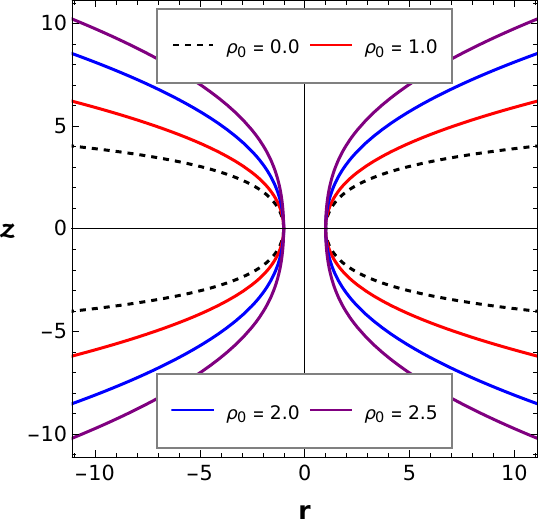}
\includegraphics[scale=0.355]{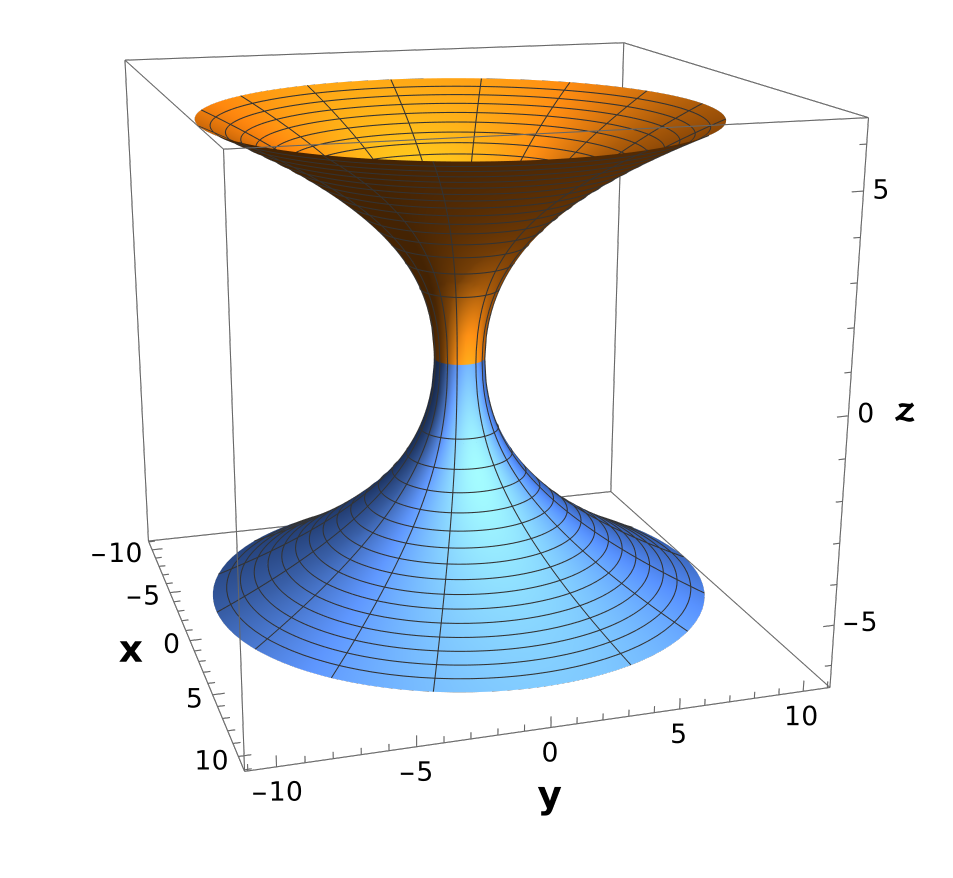}
\vspace{-0.2cm}
\caption{2D (left and middle) and 3D (right) embedding diagrams of the 
wormhole geometry. The left panel shows the effect of varying the Gauss-Bonnet 
coupling parameter $\alpha$ for fixed $\rho_0=1$, while the middle panel shows 
the effect of varying $\rho_0$ for fixed $\alpha=0.25$. The throat radius is 
fixed at $r_0=1$. The 3D embedding diagram for $\alpha=0.25$ and $\rho_0=1$ 
illustrates the two asymptotically flat regions, shown by the orange and blue 
surfaces, smoothly connected through the wormhole throat $r_0=1$.}
\label{fig6}
\end{figure}

To visualize the embedded diagram of the wormhole represented by the metric 
function~\eqref{eq22}, we consider an equitorial slice 
$\chi=\pi/2$ and $\theta=\pi/2$ at a fixed time $t$, i.e., 
at $t=\text{constant}$. This reduces the metric~\eqref{eq21} as
\begin{equation}
ds^2=\frac{dr^2}{f(r)}+ r^2 d\phi^2.
\label{eq42}
\end{equation}
In cylindrical coordinates, one can write the spacetime metric as
\begin{equation}
ds^2=dz^2+dr^2+r^2d\phi^2.
\label{eq43}
\end{equation}
In three dimensional Euclidean space the embedded surface is defined by 
$z=z(r)$. Therefore Eq.~\eqref{eq43} can be written as
\begin{equation}
ds^{2}=\left[1+\left(\frac{dz}{dr}\right)^{2}\right]dr^{2}+r^{2}d\phi^{2}.
\label{eq44}
\end{equation}
Comparing Eq.~\eqref{eq42} and~\eqref{eq44}, one can obtain
\begin{equation}
z(r)= \pm \int_{r_{0}}^{r}\!\! \sqrt{1/f(r)-1}\,\, dr.
\label{eq45}
\end{equation}
Fig.~\ref{fig6} shows the 2D and 3D embedding diagrams of the wormhole 
geometry. The left and middle panels illustrate the effects of the 
Gauss-Bonnet coupling parameter $\alpha$ and string-cloud parameter $\rho_0$, 
respectively, while the right panel shows the corresponding 3D geometry for 
$\alpha=0.25$, $\rho_0=1$, and $r_0=1$. The embedding surfaces are smooth and 
symmetric, with both branches joining continuously at the throat, confirming 
the regularity of the geometry. Increasing either $\alpha$ or $\rho_0$ causes 
the embedding curves to spread farther from the symmetry axis, indicating an 
increased spatial extent of the wormhole while keeping the throat radius 
fixed. The 3D diagram further displays the characteristic hourglass-like 
structure connecting two asymptotically flat regions, with the strongest 
bending near the throat and a gradual flattening at large distances.

\subsection{Travel Time for Traversing the Wormhole}\label{sec5sub2}

To further examine the traversability of the obtained wormhole, we estimate 
the time required for a traveler to move from one asymptotically flat region 
to the other. Following the traversability prescription proposed by Morris and 
Thorne~\cite{a3}, we assume that the traveler moves radially with a constant 
local velocity $v$, as measured by a static observer. For the metric given in 
Eq.~\eqref{eq21}, the proper time interval measured by the static observer is 
related to the coordinate time through $d\tau_s=\sqrt{f(r)}\,dt$. Using the 
definition of the local velocity, $v=d\ell/d\tau_s$, together with the proper 
radial distance given in Eq.~\eqref{eq41}, one may obtain
\begin{equation}
dt=\frac{d\ell}{v\sqrt{f(r)}}
=\frac{dr}{vf(r)}.
\label{eq47}
\end{equation}
Therefore, the coordinate time required for a traveler to move from a station 
located at $\mathcal{R}$ on one side of the wormhole to the corresponding 
station on the opposite side is
\begin{equation}
t_{\text{travel}} = 2\int_{r_0}^{\mathcal{R}}\!\! \frac{dr}{vf(r)},
\label{eq48}
\end{equation}
where $\mathcal{R}$ denotes the radial coordinate of the static station in the 
asymptotically flat region and the factor of two accounts for the two 
symmetric asymptotic regions. The proper time experienced by the traveler is 
obtained from
\begin{equation}
d\tau =
\frac{dl}{v} = \frac{dr}{v\sqrt{f(r)}},
\label{eq49}
\end{equation}
which leads to
\begin{equation}
\tau_{\text{travel}} = 2\int_{r_0}^{\mathcal{R}}\!\! \frac{dr}{v\sqrt{f(r)}}
=\frac{2\,l(\mathcal{R})}{v}.
\label{eq50}
\end{equation}
To examine the behavior of these quantities near the throat, we expand the 
metric function about $r=r_0$ as $f(r)\simeq f'(r_0)(r-r_0)$. Substituting 
this approximation into Eq.~\eqref{eq50} gives
\begin{equation}
\tau_{\text{travel}} \sim \frac{4\sqrt{r-r_0}} {v\sqrt{f'(r_0)}}.
\label{eq50a}
\end{equation}
This shows that the proper traversal time remains finite as the throat is 
approached. On the other hand, Eq.~\eqref{eq48} yields for this expansion as
\begin{equation}
 t_{\text{travel}} \sim \frac{2}{vf'(r_0)} \ln|r-r_0|,
 \label{eq50b}
\end{equation}
which diverges logarithmically in the limit $r\rightarrow r_0$. Thus, the 
coordinate traversal time diverges in the adopted static coordinate system, 
whereas the proper time measured along the traveler's worldline remains 
finite. To illustrate how the traversal properties depend on the model 
parameters, Table~\ref{tab1} presents the dimensionless proper radial 
distance, $\ell(\mathcal{R})/r_0$, together with the dimensionless coordinate 
traversal time, $v\,t_{\text{travel}}/r_0$, evaluated for a station located 
at $\mathcal{R}=10r_0$. The left side of the table shows the effect of varying 
the string cloud parameter $\rho_0$ while keeping the Gauss-Bonnet coupling 
fixed at $\alpha=0.25$, whereas the right side illustrates the influence of 
the Gauss-Bonnet coupling $\alpha$ for a fixed string cloud parameter 
$\rho_0=1$. It can be seen that both the proper radial distance and the 
coordinate traversal time increase with increasing values of either 
$\rho_0$ or $\alpha$. This indicates that the wormhole geometry becomes more 
extended with the increase of $\rho_0$ and $\alpha$, leading to a larger 
spatial separation between the throat and the observation station. At the same 
time, the increase in the coordinate traversal time reflects the stronger 
suppression of the metric function $f(r)$ in the vicinity of the throat. 
Despite this behavior, the proper traversal time remains finite through the 
relation $\tau_{\text{travel}}=2\ell(\mathcal{R})/v$, showing that the elapsed 
time measured by the traveler increases smoothly with the wormhole size.
\begin{table}[!h]
\centering
\caption{Dimensionless proper radial distance $\ell(\mathcal{R})/r_0$ and 
coordinate traversal time $v\,t_{\text{travel}}/r_0$ evaluated at 
$\mathcal{R}=10r_0$ for different values of the string cloud parameter 
$\rho_0$ and the Gauss-Bonnet coupling parameter $\alpha$.}
\vspace{5pt}
\label{tab1}
\begin{tabular}{ccc|ccc}
\hline\hline
\multicolumn{3}{c|}{$\alpha=0.25$} &
\multicolumn{3}{c}{$\rho_0=1$} \\[1pt]
\cline{1-3}\cline{4-6}
$\rho_0$ & $\ell(\mathcal{R})/r_0$ & $v\,t_{\text{travel}}/r_0$
& $\alpha$ & $\ell(\mathcal{R})/r_0$ & $v\,t_{\text{travel}}/r_0$ \\
\hline
0.0 & 10.478 & 45.901  & 0.2 & 11.294 & 128.600 \\
1.0 & 11.414 & 61.917  & 0.4 & 11.770 & 170.447 \\
2.0 & 12.878 & 104.461 & 0.6 & 12.211 & 205.742 \\
2.5 & 14.184 & 180.576 & 0.8 & 12.624 & 275.288 \\
\hline\hline
\end{tabular}
\end{table}

\subsection{Proper Acceleration}\label{sec5sub3}

A fundamental requirement for a traversable wormhole is that the proper 
acceleration experienced by a traveler remains finite throughout the journey. 
Otherwise, unrealistically large propulsion forces would be required to 
cross the throat. To examine this condition, we consider a traveler moving 
radially through the wormhole with five-velocity 
$u^\mu=\left(\dot t,\dot r,0,0,0\right)$, where an overdot denotes 
derivative with respect to the proper time $\tau$. The corresponding 
five-acceleration is defined by $a^\mu=u^\nu\nabla_\nu u^\mu$, and 
its invariant magnitude is
\begin{equation}
a=\sqrt{a_\mu a^\mu}.
\label{eq53}
\end{equation}
The explicit expressions for the nonvanishing components of $a^\mu$ and the
resulting proper acceleration are given by
\begin{align}
a^t&=\ddot t+\frac{f'(r)}{f(r)}\,\dot t\,\dot r,
\label{eq51}\\
a^r&=\ddot r+\frac12f(r)f'(r)\dot t^{\,2}
-\frac{f'(r)}{2f(r)}\dot r^{\,2},
\label{eq52}
\end{align}
which lead to
\begin{equation}
a= \sqrt{ -f(r)\left(a^t\right)^2 +\frac{\left(a^r\right)^2}{f(r)}}.
\label{eq54}
\end{equation}

To examine the traversability of the wormhole, we consider a traveler moving 
radially with a constant radial velocity $\dot r=v$, so that $\ddot r=0$. 
Applying the normalization condition $g_{\mu\nu}u^\mu u^\nu=-1$ gives
\begin{equation}
\dot t=\frac{\sqrt{f(r)+v^2}}{f(r)}.
\label{eq54a}
\end{equation}
Using Eq.~\eqref{eq54a} in Eqs.~\eqref{eq51}--\eqref{eq54}, the required 
proper acceleration becomes
\begin{equation}
a(r)=\frac{|f'(r)|}{2\sqrt{f(r)+v^2}}.
\label{eq54b}
\end{equation}
Eq.~\eqref{eq54b} gives the proper acceleration required for a traveler
to maintain a constant radial component of the five-velocity while traversing 
the wormhole.
\begin{figure}[!h]
\centering
\includegraphics[scale=0.67]{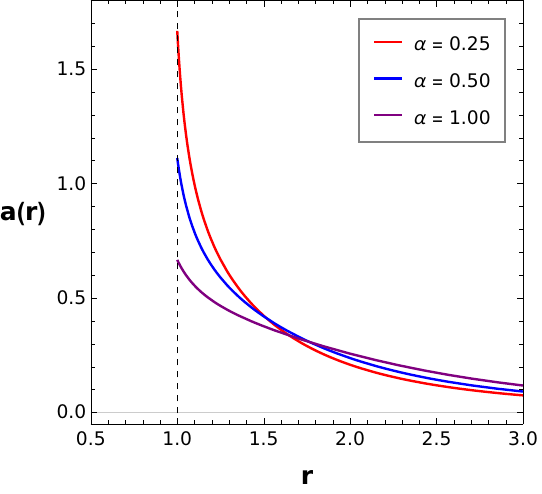}\hspace{1.0cm}
\includegraphics[scale=0.66]{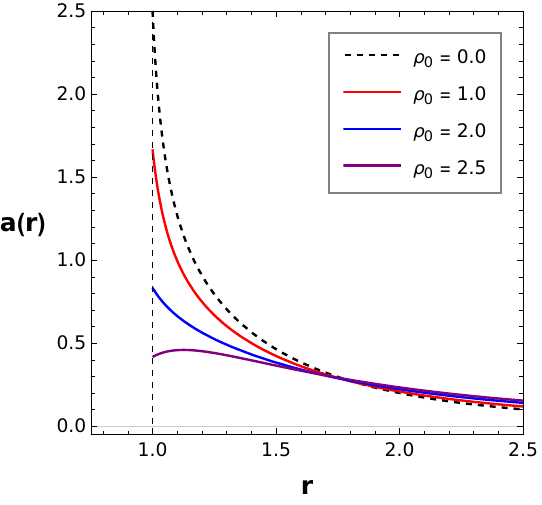}
\vspace{-0.2cm}
\caption{Proper acceleration $a(r)$ as a function of the radial
coordinate $r$ for different values of the Gauss-Bonnet coupling
parameter $\alpha$ with fixed string-cloud density parameter $\rho_0=1$ (left
panel) and for different values of the string-cloud density $\rho_0$
with fixed Gauss-Bonnet coupling $\alpha=0.25$ (right panel). In both
panels, the throat radius is fixed at $r_0=1$ and the traveler is
assumed to move with a constant radial component of the five-velocity
$v=0.2$. The vertical dashed line marks the location of the wormhole
throat.}
\label{fig6a}
\end{figure}
Fig.~\ref{fig6a} shows the proper acceleration required for radial traversal. 
The acceleration is strongest near the throat and decreases rapidly with 
distance. Increasing the Gauss-Bonnet coupling parameter $\alpha$ lowers the 
near-throat acceleration, while the effect of the string-cloud parameter 
$\rho_0$ is mainly confined to the throat region and becomes negligible far 
away. The locally measured velocity is different from the radial five-velocity 
$v$ and is given by
\begin{equation}
v_{\text{local}}=\frac{1}{f(r)}\frac{dr}{dt}
=\frac{v}{\sqrt{f(r)+v^2}},
\label{eq54c}
\end{equation}
with $0\leq v_{\text{local}}\leq1$. Thus, $v$ itself is not the physical speed 
measured by a static observer. At the throat, $v_{\text{local}}\to1$ for 
outward motion, while at large distances, where $f(r)\to1$, it approaches a 
value below the speed of light. Hence, the traveler remains subluminal 
throughout the journey.

For a freely falling traveler, the motion follows a geodesic satisfying
$u^\nu\nabla_\nu u^\mu=0$. Consequently, $a^\mu=0$ and $a=0$ showing that no 
proper acceleration is experienced during the traversal. Therefore, an 
inertial traveler can cross the wormhole without requiring any propulsion, and 
no divergent acceleration is encountered at the throat.

\subsection{Radial and Lateral Tidal Accelerations}\label{sec5sub4}

Although a traveler following a geodesic experiences vanishing proper 
acceleration, finite tidal forces may still arise due to the curvature of 
spacetime. These forces describe the relative acceleration between neighboring 
freely falling particles and determine whether an extended object can safely 
traverse the wormhole without suffering excessive stretching or compression. 
We therefore examine the radial and lateral tidal accelerations associated 
with the present wormhole geometry. The relative acceleration between 
neighboring geodesics is governed by the geodesic deviation equation,
\begin{equation}
\frac{D^{2}\xi^{\hat{\mu}}}{d\tau^{2}}
=-R^{\hat{\mu}}{}_{\hat{\nu}\hat{\rho}\hat{\sigma}}u^{\hat{\nu}} u^{\hat{\rho}} \xi^{\hat{\sigma}},
\label{eq55}
\end{equation}
where $D = d x^\mu \nabla_\mu$ is the directional covariant deviation and 
$\xi^{\hat{\mu}}$ denotes the infinitesimal separation vector between 
neighboring particles and $u^{\hat{\mu}}$ is the five-velocity measured in the 
traveler's local orthonormal frame. By the equivalence principle, a freely 
falling observer can always construct an instantaneous locally inertial frame 
in which the traveler is momentarily at rest. In this comoving frame, 
$u^{\hat{\mu}}=(1,0,0,0,0)$ and therefore the geodesic deviation equation 
reduces to
\begin{equation}
\frac{D^{2}\xi^{\hat{i}}}{d\tau^{2}}=-R^{\hat{i}}{}_{\hat0\hat{j}\hat0}\,\xi^{\hat{j}},
\label{eq56}
\end{equation}
where $\hat{i}$ and $\hat{j}$ denote spatial orthonormal indices. The required 
orthonormal curvature components are obtained by transforming the 
coordinate-basis Riemann tensor into the traveler's instantaneous rest frame. 
The resulting radial and lateral tidal accelerations are
\begin{equation}
\Delta a_{\text{radial}} = -\frac{f''(r)}{2}\,\xi,
\qquad
\Delta a_{\text{lateral}}= -\frac{f'(r)}{2r}\,\xi,
\label{eq57}
\end{equation}
where $\xi$ denotes the characteristic size of the traveler. The sign of the 
tidal acceleration indicates whether neighboring geodesics undergo relative 
stretching (positive) or compression (negative), whereas its magnitude 
determines the strength of the tidal force. Using metric 
function~\eqref{eq22}, the radial and lateral tidal accelerations at the 
throat become
\begin{equation}
\Delta a_{\text{radial}}(r_{0})
=\frac{27r_{0}^{4}-12r_{0}^{3}\rho_0-8\alpha\left(18\alpha+\rho_0^{2}\right)}
{9\left(r_{0}^{2}+4\alpha\right)^{3}}\,\xi,
\qquad
\Delta a_{\text{lateral}}(r_{0})
=- \frac{6r_{0}-2\rho_0}{6r_{0}(r_{0}^{2}+4\alpha)}\,\xi.
\label{eq58}
\end{equation}
In particular, both tidal accelerations remain finite at the throat provided 
that $f'(r_{0})$ and $f''(r_{0})$ are finite, which is satisfied for the 
present wormhole solution. Hence, although a freely falling traveler 
experiences zero proper acceleration, the curvature of spacetime generates 
finite tidal forces. The above expressions, therefore, quantify the radial and 
lateral tidal forces experienced by a freely falling traveler and provide the 
basis for assessing the traversability of the proposed 5D EGB wormhole.

The tidal accelerations appearing in Eqs.~\eqref{eq58} are expressed in the 
geometrized units adopted throughout this work. A quantitative comparison with 
human tolerability limits requires specifying the physical length scale of 
the wormhole geometry and restoring the appropriate conversion factors to SI 
units. Thus, to recover the corresponding physical tidal accelerations, let the 
physical radial coordinate be written as $r_{\text{phys}}=\mathcal{R}_{0}\,r$, 
where $\mathcal{R}_{0}$ denotes the characteristic length scale of the 
wormhole, which may be identified with the physical throat radius. 

Moreover, restoring SI units requires the introduction of an overall 
$c^{2}$ factor in the tidal accelerations, which converts the geometrized 
accelerations into the physical accelerations measured in ms$^{-2}$. 
Consequently, the radial and the lateral tidal accelerations in SI units take 
the form:
\begin{equation}
\Delta a_{\text{radial}}^{\text{(SI)}}
=-\frac{c^{2}}{2R_{0}^{2}}\,f''(r)\,\xi,
\qquad
\Delta a_{\text{lateral}}^{\text{(SI)}}
=-\frac{c^{2}}{2R_{0}^{2}}\frac{f'(r)}{r}\,\xi.
\label{eq61}
\end{equation}
where $\xi$ denotes the physical size of the traveler. Evaluating these 
expressions at the throat, the physical tidal accelerations at the throat 
become
\begin{equation}
\Delta a_{\text{radial}}^{\text{(SI)}}(r_{0})
=\frac{c^{2}\xi}{9R_{0}^{2}}\,\frac{27r_{0}^{4}-12r_{0}^{3}\rho_0-8\alpha(18\alpha+\rho_0^{2})}
{(r_{0}^{2}+4\alpha)^{3}},
\qquad
\Delta a_{\text{lateral}}^{\text{(SI)}}(r_{0})
=
-\frac{c^{2}\xi}{6R_{0}^{2}}\,\frac{6r_{0}-2\rho_0}
{r_{0}(r_{0}^{2}+4\alpha)}.
\label{eq62}
\end{equation}
These expressions show that both the radial and lateral tidal accelerations 
scale inversely with the square of the physical throat size, i.e., 
$\Delta a \propto 1/R_{0}^{2}$. Therefore, for fixed values of the 
dimensionless parameters $(\alpha,\rho_0,r_{0})$, increasing the physical size 
of the wormhole rapidly suppresses the tidal forces experienced by a traveler. 
Conversely, sufficiently small wormholes can produce extremely large tidal 
accelerations even when the geometry is perfectly regular. To estimate the 
minimum physical size required for comfortable human traversal, we require 
that both the radial and lateral tidal accelerations do not exceed the 
Earth's gravitational acceleration, i.e.,
\begin{equation}
|\Delta a_{\text{radial}}^{(\text{SI})}|\leq g_{\oplus},
\qquad
|\Delta a_{\text{lateral}}^{(\text{SI})}|\leq g_{\oplus},
\label{eq63}
\end{equation}
where $g_{\oplus} \simeq 9.80665~\text{m\,s}^{-2}$ is the Earth's gravitational 
acceleration. These radial and lateral constraints imply
\begin{equation}
R_0^{\text{radial}} \geq c \sqrt{\frac{\xi}{9g_{\oplus}}\left|\frac{27r_0^4-12r_0^3\rho_0-8\alpha(18\alpha+\rho_0^2)}{(r_0^2+4\alpha)^3}\right|},
\qquad
R_0^{\text{lateral}} \geq c \sqrt{\frac{\xi}{6g_{\oplus}}\left|\frac{6r_0-2\rho_0}{r_0(r_0^2+4\alpha)}\right|}.
\label{eq64}
\end{equation}
Therefore, the physical throat radius must satisfy $R_0 \geq R_0^{\text{min}} = 
\max\left(R_0^{\text{radial}},\,R_0^{\text{lateral}}\right)$, which defines the 
minimum throat radius required for a traveler of characteristic size $\xi$ to 
experience tidal accelerations no greater than the Earth's surface gravity. 
Throughout this analysis, we adopt a representative traveler size of 
$\xi=2~\text{m}$. The corresponding values of $R_0^{\text{min}}$ for different 
choices of the Gauss-Bonnet coupling parameter $\alpha$ and the string cloud 
parameter $\rho_0$ are listed in Table~\ref{tab2}. For a fixed value of 
$\rho_0$, the minimum allowable throat radius decreases as the Gauss-Bonnet 
coupling $\alpha$ increases. This indicates that the higher-curvature 
corrections associated with the Gauss-Bonnet term effectively reduce the 
tidal forces experienced by the traveler, allowing safe traversal through 
comparatively smaller wormholes. On the other hand, for a fixed value of 
$\alpha$, the dependence of $R_0^{\text{min}}$ on the string cloud parameter is 
non-monotonic. As $\rho_0$ increases from zero, the required throat radius 
initially decreases, reaches a minimum near $\rho_0\simeq2$, and then 
increases for larger values of $\rho_0$. This behavior suggests that there 
exists an optimal string cloud density for which the tidal constraints are 
minimized, thereby providing the most favorable conditions for traversal. The 
results presented in Table~\ref{tab2} further indicate that the minimum throat 
radius required to keep the tidal accelerations within humanly tolerable 
limits is typically of the order of $10^{5}\,\text{km}$. Therefore, although 
the obtained wormhole solution satisfies the geometric requirements for 
traversability and the tidal accelerations remain finite, comfortable human 
traversal would require an astrophysically large throat. Nevertheless, the 
inclusion of the Gauss-Bonnet term substantially relaxes this requirement 
by reducing the minimum throat radius needed to satisfy the tidal constraints.
\begin{table}[!h]
\centering
\caption{Minimum physical throat radius $R_0^{\text{min}}$ (in km) required for 
comfortable human traversal, assuming a traveler of characteristic size 
$\xi=2~\text{m}$ and imposing the condition that the tidal accelerations do 
not exceed the Earth's surface gravity.}
\vspace{5pt}
\label{tab2}
\begin{tabular}{c|cccccc}
\hline\hline
$\alpha$ & $\rho_0=0$ & $\rho_0=0.5$ & $\rho_0=1.0$ &
$\rho_0=1.5$ & $\rho_0=2.0$ & $\rho_0=2.5$ \\
\hline
0.1 & 137734 & 119871 & 97317 & 80909 & 66062 &{\text{min}} 83704 \\
0.2 & 100911 & 92119 & 82394 & 71355 & 58261 & 80940 \\
0.3 & 91278 & 83325 & 74528 & 64543 & 61165 & 76952 \\
0.4 & 83963 & 76648 & 68556 & 59371 & 61687 & 73040 \\
0.5 & 78165 & 71355 & 63822 & 55271 & 60795 & 69480 \\
\hline
\hline
\end{tabular}
\end{table}

%
The analyses presented in the above subsections provide a comprehensive 
assessment of the traversability properties of the obtained 5D EGB 
wormhole~\cite{a3,a31}. The proper radial distance remains finite and increases smoothly away from the throat, while the embedding diagrams confirm that the 
two asymptotically flat regions are connected through a regular geometric 
bridge. Together with the finite curvature invariants discussed in 
Sec.~\ref{sec3}, these results demonstrate that the wormhole geometry is free 
from curvature singularities.

The traversal analysis further shows that the proper time experienced by a 
traveler remains finite, whereas the corresponding coordinate time diverges 
logarithmically in the adopted static coordinate system as the throat is 
approached. The proper radial distance and traversal time both increase with 
the Gauss-Bonnet coupling parameter and the string cloud parameter, indicating 
that these quantities effectively increase the radial extent of the wormhole 
geometry.

The proper acceleration analysis reveals that a freely falling traveler 
follows a geodesic trajectory with vanishing five-acceleration, so that no 
external propulsion is required during the journey. The remaining mechanical 
effects arise from the tidal forces associated with the spacetime curvature. 
These tidal accelerations remain finite throughout the spacetime and decrease 
with increasing physical throat radius. For the representative parameter 
values considered in this work, the tidal constraints can be satisfied when 
the physical throat radius is of the order of $10^{5}\,\text{km}$, with the 
precise value depending on the Gauss-Bonnet coupling and the string cloud 
parameter.

Further, the asymptotically flat behavior of the metric ensures that the 
spacetime approaches the Minkowski limit far from the throat, allowing 
observers to be located in regions where gravitational effects become 
negligible. Taken together, the regularity of the geometry, the finite proper 
traversal time, the absence of proper acceleration for freely falling 
observers and the bounded tidal forces indicate that the obtained solution 
possesses the essential features expected of a physically traversable 
wormhole~\cite{a57,a58}. Moreover, the physical string cloud satisfies the 
classical energy conditions, while the higher-curvature Gauss-Bonnet sector 
provides the effective gravitational contribution required to sustain the 
wormhole geometry.

\section{Null Geodesics and Optical Signatures}\label{sec6}

We now investigate the propagation of light in this wormhole geometry. The 
study of null geodesics provides a direct connection between the spacetime 
structure and potentially observable phenomena, such as gravitational lensing, 
photon capture and the shadow cast by the wormhole~\cite{a36,a37,a38,a59}.

\subsection{Null Geodesic Equations and Photon-Sphere}\label{sec6sub1}

\begin{figure}[!h]
\centering
\includegraphics[scale=0.58]{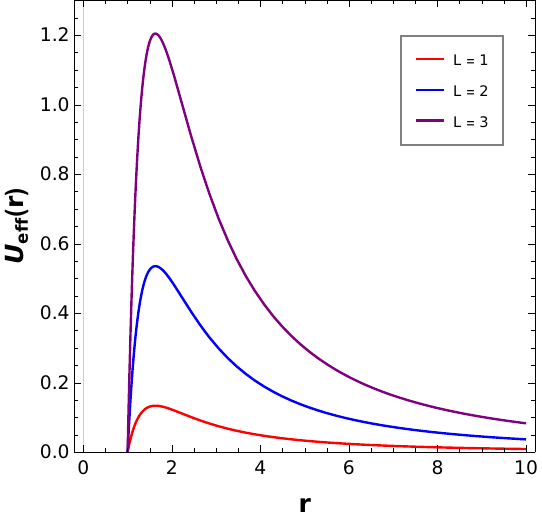}\hspace{0.5cm}
\includegraphics[scale=0.595]{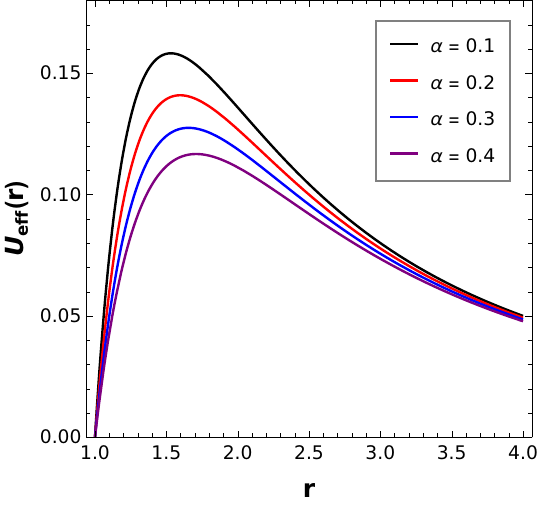}\hspace{0.5cm}
\includegraphics[scale=0.6]{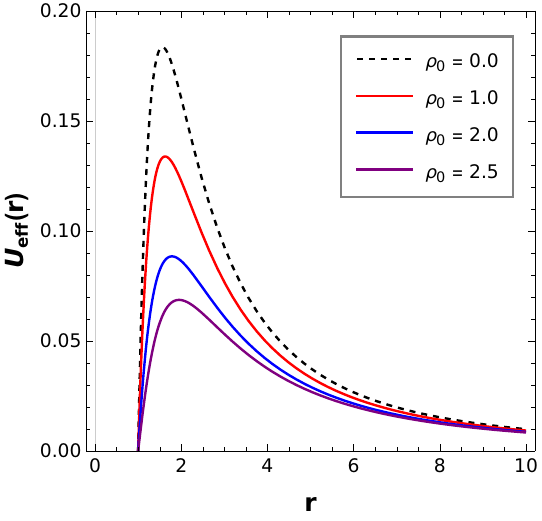}
\vspace{-0.2cm}
\caption{Effective potential $U_{\text{eff}}(r)$ for different values of the 
angular momentum $L$ (left panel), the Gauss-Bonnet coupling parameter 
$\alpha$ (centre panel), and the string cloud parameter $\rho_0$ 
(right panel). Unless otherwise stated, the parameters are fixed as 
$r_0=1$, $\alpha=0.25$, $\rho_0=1$, and $L=1$ in the concerned plots.}
\label{fig7}
\end{figure}

The propagation of light in the wormhole spacetime is governed by null 
geodesics, which satisfy the condition $ds^{2}=0$. In general, owing to the 
spherical symmetry of spacetime, the motion of any particle can always be 
confined to an invariant equatorial hypersurface without loss of generality. We 
therefore choose $\chi=\theta=\pi/2$ such that the metric~\eqref{eq21} 
reduces to
\begin{equation}
ds^{2} = -f(r)\,dt^{2}+\frac{dr^{2}}{f(r)}+r^{2}d\phi^{2}.
\label{eq65}
\end{equation}
Thus, the geodesic motion may be derived from the Lagrangian
\begin{equation}
\mathcal{L}=\frac{1}{2}\left[-f(r)\dot{t}^{2}+\frac{\dot{r}^{2}}{f(r)}+r^{2}\dot{\phi}^{2}\right],
\label{eq66}
\end{equation}
where an overdot denotes a differentiation with respect to the affine parameter $\tau$. Since the metric is independent of the coordinates $t$ and $\phi$, there exist two conserved quantities corresponding to the timelike and rotational Killing vectors. These conserved respective energy and angular momentum of the photon are given by
\begin{equation}
\frac{\partial \mathcal{L}}{\partial \dot{t}}
=
-f(r)\dot{t}
\equiv
-E,
\qquad
\frac{\partial \mathcal{L}}{\partial \dot{\phi}}
=
r^{2}\dot{\phi}
\equiv
L.
\label{eq67}
\end{equation}
Consequently,
\begin{equation}
\dot{t}
=
\frac{E}{f(r)},
\qquad
\dot{\phi}
=
\frac{L}{r^{2}}.
\label{eq68}
\end{equation}
Now, using the condition for null geodesics ($\mathcal{L}=0$) and substituting 
$\dot{t}$ and $\dot{\phi}$ from Eq.~\eqref{eq68} in Eq.~\eqref{eq66}, one 
may obtain
\begin{equation}
-\frac{E^{2}}{f(r)}
+\frac{\dot{r}^{2}}{f(r)}
+\frac{L^{2}}{r^{2}}
=0.
\label{eq69}
\end{equation}
Therefore, the radial equation of motion of photons becomes
\begin{equation}
\dot{r}^{2}=E^{2}-\frac{L^{2}}{r^{2}}f(r)=E^{2}-U_{\text{eff}}(r),
\label{eq70}
\end{equation}
where
\begin{equation}
U_{\text{eff}}(r)=\frac{L^{2}}{r^{2}}f(r)
\label{eq71}
\end{equation}
represents an effective potential barrier that governs the radial motion of 
photons. It is convenient to introduce the impact parameter $b=L/E$ in terms of which Eq.~\eqref{eq70} can be written as
\begin{equation}
\dot{r}^{2}=E^{2}\left[1-\frac{b^{2}}{r^{2}}f(r)\right].
\label{eq72}
\end{equation}
Furthermore, using
$dr/d\phi
=
\dot{r}/\dot{\phi}
=
r^{2} \dot{r}/L$,
the trajectory equation for photon motion becomes
\begin{equation}
\left(\frac{dr}{d\phi}\right)^{2}
=
r^{4}
\left[
\frac{1}{b^{2}}
-
\frac{f(r)}{r^{2}}
\right].
\label{eq74}
\end{equation}
Eqs.~\eqref{eq70}, \eqref{eq72} and \eqref{eq74} completely characterize the 
null geodesics of the present 5D EGB wormhole spacetime. In particular, 
Eq.~\eqref{eq70} admits the form of an energy conservation equation. 
Fig.~\ref{fig7} illustrates the effective potential given by Eq.~\eqref{eq71} 
governing the motion of photons in the 5D EGB wormhole spacetime. For all 
parameter choices, the potential vanishes at the throat, increases to a single 
maximum, and gradually approaches zero at large radial distances. The presence 
of a unique maximum corresponds to an unstable circular photon orbit, which 
defines the photon-sphere of the wormhole. The left panel shows the influence 
of the photon angular momentum $L$. As expected, increasing $L$ raises the 
height of the effective potential while leaving its overall profile 
essentially unchanged. This behaviour follows directly from the 
proportionality $U_{\text{eff}}\propto L^{2}$ and reflects the stronger 
centrifugal barrier experienced by photons with larger angular momentum. The 
middle panel demonstrates the effect of the Gauss-Bonnet coupling parameter 
$\alpha$. Increasing $\alpha$ lowers the maximum of the effective potential, 
while the location of the peak changes only slightly. This indicates that 
higher-curvature corrections primarily reduce the strength of the 
gravitational potential governing photon motion, leading to weaker confinement 
of null geodesics near the throat. The right panel illustrates the role of the 
string cloud parameter $\rho_0$. Similar to the Gauss-Bonnet coupling, 
increasing $\rho_0$ suppresses the effective potential, with the vacuum 
configuration ($\rho_0=0$) exhibiting the highest potential barrier. This 
behaviour suggests that the presence of the string cloud further weakens the 
gravitational confinement of photons in the vicinity of the throat.

Overall, both the Gauss-Bonnet coupling and the string cloud parameter 
significantly influence the effective potential and, consequently, the 
properties of the photon-sphere and the optical appearance of the wormhole. 
These modifications are expected to affect observable phenomena associated 
with null geodesics, including photon capture and the size of the wormhole 
shadow.

\begin{figure}[!h]
\centering
\includegraphics[scale=0.57]{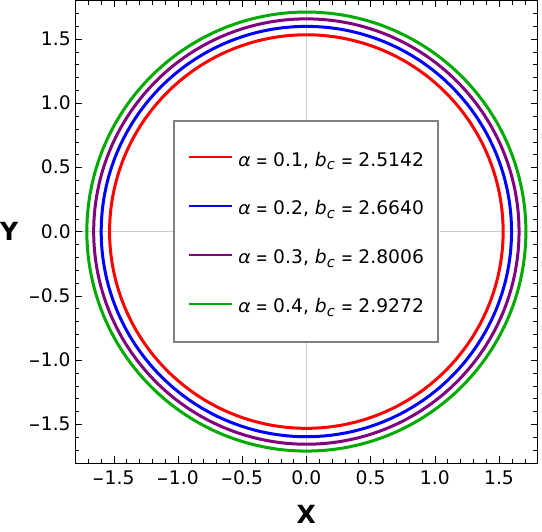}\hspace{0.5cm}
\includegraphics[scale=0.55]{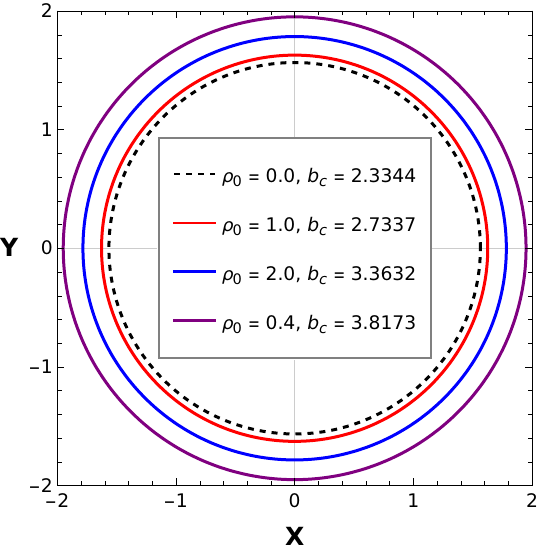}\hspace{0.5cm}
\includegraphics[scale=0.55]{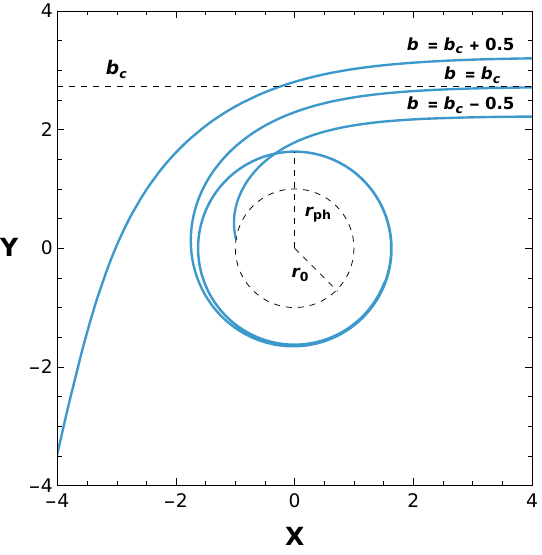}
\vspace{-0.2cm}
\caption{Photon-sphere of the wormhole for different values of the 
Gauss-Bonnet coupling parameter $\alpha$ (left) and string-cloud parameter 
$\rho_0$ (middle), showing the associated values of the critical impact 
parameter $b_c$. The right panel shows the representative photon trajectories 
for $r_0=1$, $\alpha=0.25$ and $\rho_0=1$. The dashed circle marks the 
wormhole throat, while the horizontal dashed line indicates the critical 
impact parameter $b_c$, separating scattered and transmitted photon 
trajectories.}
\label{fig8}
\end{figure}

%
The photon-sphere is formed by unstable circular null geodesics and plays a 
central role in determining the optical appearance of the 
wormhole~\cite{a38,a59,a60}. These circular orbits correspond to the extrema 
of the effective potential. From Eq.~\eqref{eq70}, the conditions for a 
circular photon orbit are $\dot r=0$ and $dU_{\text{eff}}/dr=0$, while the 
additional requirement $d^{2}U_{\text{eff}}(r_{\text{ph}})/dr^{2}<0$ ensures 
that the orbit is unstable. Using the effective potential given in Eq.~\eqref{eq71}, the photon-sphere radius $r_{\text{ph}}\ge r_0$ is determined from
\begin{equation}
r_{\text{ph}}f'(r_{\text{ph}}) - 2f(r_{\text{ph}})=0,
\label{eq75}
\end{equation}
while the corresponding critical impact parameter is
\begin{equation}
b_c=\frac{r_{\text{ph}}}{\sqrt{f(r_{\text{ph}})}}.
\label{eq76}
\end{equation}
The value of $b_c$ separates photons with different orbital behaviours. 
Photons with $b>b_c$ are deflected by the gravitational field and eventually 
return to the asymptotic region, whereas those with $b<b_c$ overcome the 
potential barrier and propagate toward the wormhole throat. The critical 
trajectory corresponding to $b=b_c$ asymptotically approaches the unstable 
circular orbit at the photon-sphere and therefore forms the boundary between 
scattered and transmitted null geodesics. To visualize the photon 
trajectories, the numerically integrated null geodesics are projected onto 
the equatorial plane using the Cartesian coordinates~\cite{a61}
\begin{equation}
X=r(\tau)\cos\phi(\tau),
\qquad
Y=r(\tau)\sin\phi(\tau),
\label{eq77}
\end{equation}
where $r(\tau)$ and $\phi(\tau)$ are obtained by numerically solving the null 
geodesic equations. These coordinates are introduced solely for the purpose of 
ray-tracing visualization. The left and middle panels of 
Fig.~\ref{fig8} show how the photon-sphere radius and critical impact 
parameter $b_c$ vary with the Gauss-Bonnet coupling parameter $\alpha$ and 
string-cloud parameter $\rho_0$. Both quantities increase as $\alpha$ or 
$\rho_0$ increases, while the photon-ring remains circular due to the 
spherical symmetry of the spacetime. The right panel illustrates the three 
types of photon trajectories for $\alpha=0.25$, $\rho_0=1$, and $r_0=1$. 
Photons with $b>b_c$ are scattered back to infinity, whereas those with 
$b<b_c$ pass through the throat into the other asymptotically flat region. 
For $b=b_c$, the photon approaches the unstable circular orbit at 
$r_{\rm ph}$, marking the boundary between the scattered and transmitted 
trajectories and determining the shadow size.

\subsection{Shadows}\label{sec6sub2}

The shadow of a compact object appears as a dark region in the observer's sky, 
produced by photons that asymptotically spiral toward unstable circular null 
orbits instead of reaching the observer. In asymptotically flat spacetime, 
the boundary of the shadow is determined by the critical photon trajectories 
that separate escaping photons from those captured by the unstable photon 
orbit. For a static, spherically symmetric spacetime, the shadow radius 
measured by an observer at infinity is therefore given by the critical impact 
parameter. Using Eq.~\eqref{eq76}, the corresponding shadow radius can be 
written as~\cite{a35a}
\begin{equation}
R_{\text{sh}} = \frac{r_{\text{ph}}}{\sqrt{f(r_{\text{ph}})}}.
\label{eq78}
\end{equation}

Following the standard celestial-coordinate formalism for 5D spacetimes and 
adapting it to the angular sector $d\Omega_3^2$, the stereographic projection 
of the shadow onto the observer's image plane through the celestial 
coordinates $(X,Y)$ can determine the shadow's apparent shape, where the 
coordinates $(X,Y)$ are defined as~\cite{a62,a63}
\begin{equation}
X = \lim_{r_\text{ob}\to\infty}
\left[r_\text{ob}^{2}\left(\sin\chi_\text{ob}\frac{d\theta}{dr}\bigg|_{(r_\text{ob},\theta_\text{ob})}\!\!\!\!+\cos\chi_\text{ob}\frac{d\phi}{dr}\bigg|_{r_\text{ob}}
\right)\right],
\qquad
Y = \lim_{r_\text{ob}\to\infty}
\left(r_\text{ob}^{2}\sin\chi_\text{ob}\frac{d\chi}{dr}\bigg|_{\chi_\text{ob}}\right),
\label{eq79}
\end{equation}
where $r_\text{ob}$ denotes the radial position of the observer, while
$\theta_\text{ob}$ and $\chi_\text{ob}$ specify its angular position
on the three-sphere. For an equatorial observer 
($\theta_\text{ob}=\chi_\text{ob}=\pi/2$), the shadow is circular owing to the 
spherical symmetry of the spacetime and its boundary is described by 
$X^{2}+Y^{2}=R_{\text{sh}}^{2}$.

\begin{figure}[!h]
\centering
\includegraphics[scale=0.6]{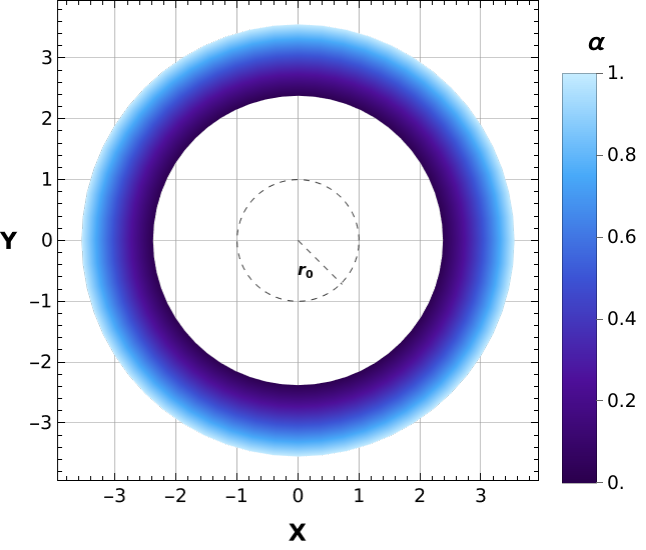}\hspace{1cm}
\includegraphics[scale=0.6]{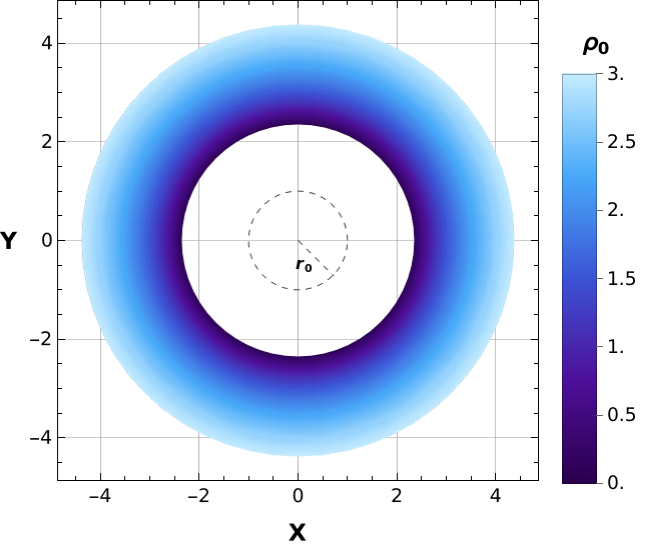}
\vspace{-0.2cm}
\caption{Stereographic mapping of the shadows of the wormhole with the 
Gauss-Bonnet coupling parameter $\alpha$ for fixed $\rho_0=1$ and $r_{0}=1$ 
(left) and with the string-cloud density parameter $\rho_0$ for fixed 
$\alpha=0.25$ and $r_{0}=1$ (right). The dashed circle denotes the wormhole 
throat at $r=r_{0}$.}
\label{fig9}
\end{figure}

Fig.~\ref{fig9} presents the shadows of the 5D EGB wormhole for different 
choices of the model parameters. The left panel illustrates the influence of 
the Gauss-Bonnet coupling parameter $\alpha$ while keeping $\rho_0=1$ fixed. 
It can be seen that the shadow radius increases gradually with increasing 
$\alpha$. This indicates that the unstable circular photon orbit shifts to 
larger radii as the Gauss-Bonnet contribution becomes stronger, leading to a 
larger apparent shadow for a distant observer. The right panel shows the 
variation of the shadow with the string-cloud density parameter $\rho_0$ for 
a fixed value of $\alpha=0.25$. A similar trend is observed, where larger 
values of $\rho_0$ produce a larger shadow. This occurs because the 
string-cloud matter modifies the spacetime geometry encountered by photons, 
causing the photon-sphere to move outward and increasing the corresponding 
critical impact parameter. It is also evident from both panels that the shadow 
retains a perfectly circular shape for all parameter values considered. This 
is expected since the wormhole spacetime is static and spherically symmetric. 
As a result, varying $\alpha$ and $\rho_0$ changes only the size of the shadow 
through their influence on the location of the photon-sphere, while its 
overall shape remains unaffected. These behaviours of shadows of the wormhole
in terms of parameters $\alpha$ and $\rho_0$ are in accordance with the
corresponding behaviours of photon-spheres as discussed above. 

\subsection{Global Causal Structure}\label{sec6sub3}

To examine the global causal structure of the wormhole spacetime~\cite{a64}, 
it is convenient to introduce the proper radial coordinate $\ell$, defined by 
Eq.~\eqref{eq41}. Unlike the areal radius $r$, which attains its minimum value 
at the throat $r=r_0$, the proper radial coordinate provides a regular 
description of the geometry across the throat as seen in Sec.~\ref{sec5sub1}. 
By definition, the throat is located at $\ell=0$, whereas the two 
asymptotically flat regions are represented by $\ell\rightarrow\pm\infty$. 
Thus, positive and negative values of $\ell$ correspond to the two sides of 
the wormhole, allowing the entire spacetime to be described using a single 
continuous radial coordinate. We introduce an auxiliary radial coordinate 
$x(\ell)$ defined by
\begin{equation}
\frac{dx}{d\ell}=\frac{1}{\sqrt{\mathcal{F}(\ell)}},
\label{eq79a}
\end{equation}
where $\mathcal{F}(\ell)\equiv f[r(\ell)]$, with $r(\ell)$ obtained by 
numerically inverting the proper radial distance relation~\eqref{eq41}. We 
solve Eq.~\eqref{eq79a} numerically with $x(\ell_{\min})=0$, taking 
$\ell_{\min}$ close to the throat. Since $x$ is defined up to an additive 
constant, we freely shift its origin to simplify the construction of the 
compactified coordinates. Therefore, we define the null 
coordinates~\cite{a65}
\begin{equation}
u=t-x(\ell),
\qquad
v=t+x(\ell),
\label{eq80}
\end{equation}
and the Kruskal-type coordinates
\begin{equation}
U=-e^{-\kappa u},
\qquad
V=e^{\kappa v},
\label{eq81}
\end{equation}
where $\kappa=f'(r_0)/2$ is the surface gravity. The spacetime is then 
conformally compactified according to
\begin{equation}
T=\tan^{-1}(V)+\tan^{-1}(U),
\qquad
X=\tan^{-1}(V)-\tan^{-1}(U).
\label{eq82}
\end{equation}
To obtain the maximal extension across the throat, the compactified 
coordinates were extended from the region $\ell>0$ to the entire domain 
$\ell\in(-\infty,\infty)$ by imposing reflection symmetry about the throat 
located at $l=0$. The extension is defined by
\begin{equation}
\mathcal{T}(t,\ell)=T(t,|\ell|),
\qquad
\mathcal{X}(t,\ell)=
\begin{cases}
\;\;X(t,\ell), & \ell\ge 0,\\
-X(t,|\ell|), & \ell<0,
\end{cases}
\label{eq83}
\end{equation}
which implies
\begin{equation}
\mathcal{T}(t,\ell)=\mathcal{T}(t,-\ell),
\qquad
\mathcal{X}(t,\ell)=-\mathcal{X}(t,-\ell).
\label{eq84}
\end{equation}
The compactified spacetime exhibits an exact reflection symmetry about the 
wormhole throat. As a result, the proper radial coordinate covers the entire 
range $\ell\in(-\infty,\infty)$, with the throat located at $\ell=0$ and the 
two asymptotically flat regions lying at $\ell\rightarrow\pm\infty$. This 
construction provides a smooth extension of the spacetime across the throat, 
allowing both Universes connected by the wormhole to be represented within a 
single conformal diagram.

\begin{figure}[!h]
\centering
\includegraphics[scale=0.8]{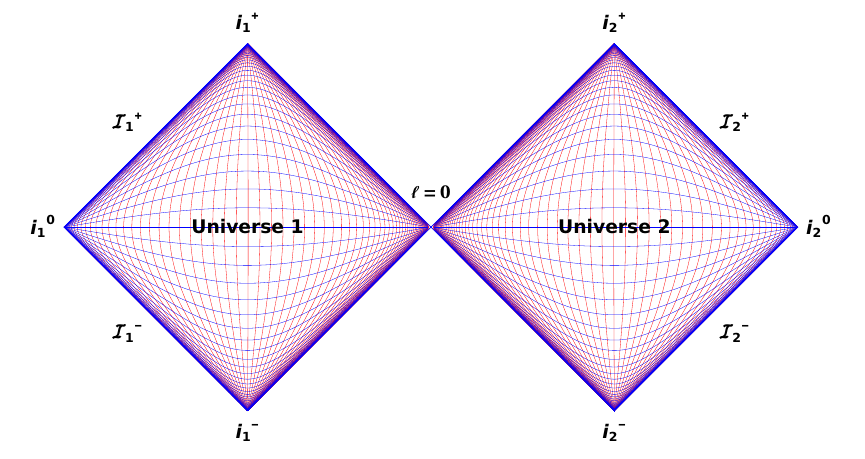}
\vspace{-0.2cm}
\caption{Penrose diagram of the 5D EGB wormhole spacetime obtained using the 
proper radial coordinate extension. The throat, located at $\ell=0$ (or 
equivalently $r=r_0$), smoothly connects two asymptotically flat regions 
labelled Universe~1 and Universe~2. The red curves represent hypersurfaces of 
constant proper radial coordinate $\ell$, while the blue curves denote 
hypersurfaces of constant coordinate time $t$. The points $i_{1,2}^{\pm}$ 
and $i_{1,2}^{0}$ correspond to future/past timelike infinities and spatial 
infinities, respectively, whereas $\mathcal{I}_{1,2}^{\pm}$ denote future 
and past null infinities.}
\label{fig10}
\end{figure}
The corresponding Penrose diagram is shown in Fig.~\ref{fig10}. Due to the 
reflection symmetry described by Eq.~\eqref{eq84}, the spacetime consists of 
two identical asymptotically flat regions joined by a regular throat at 
$\ell=0$ (or equivalently $r=r_0$). The left and right portions of the diagram 
represent the regions $\ell<0$ and $\ell>0$, respectively. The conformal 
boundaries 
follow directly from the asymptotic behaviour of the compactified coordinates 
$(\mathcal{T},\mathcal{X})$. For fixed time, the limit $\ell\rightarrow+\infty$ 
corresponds to $\mathcal{X}\rightarrow+\pi$, identifying the spatial infinity 
$i_2^{0}$ of the right asymptotic region. Likewise, $\ell\rightarrow-\infty$ 
gives $\mathcal{X}\rightarrow-\pi$, corresponding to the spatial infinity 
$i_1^{0}$ of the left asymptotic region. The timelike infinities are obtained 
by taking the limits $t\rightarrow\pm\infty$ while keeping $\ell$ fixed. The 
upper vertices of the diagram represent the future timelike infinities 
$i_1^{+}$ and $i_2^{+}$, whereas the lower vertices correspond to the past 
timelike infinities $i_1^{-}$ and $i_2^{-}$. Similarly, future null infinity 
$\mathcal{I}^{+}$ is reached in the limit $v\rightarrow+\infty$ with $u$ held 
fixed, while past null infinity $\mathcal{I}^{-}$ corresponds to 
$u\rightarrow-\infty$ with finite $v$. These null boundaries form the diagonal 
edges of the Penrose diagram and are denoted by $\mathcal{I}_{1}^{\pm}$ and 
$\mathcal{I}_{2}^{\pm}$. The causal structure is further illustrated by the 
radial null geodesics. Since they satisfy either $u=\text{constant}$ or 
$v=\text{constant}$, Eq.~\eqref{eq81} implies that $U$ or $V$ remains 
constant, respectively. Using Eq.~\eqref{eq82}, one finds $dT=dX$ for outgoing 
null rays and $dT=-dX$ for ingoing null rays. Consequently, radial light rays 
are represented by straight lines with slopes $\pm1$, appearing at an angle 
of $45^\circ$ in the compactified $(T,X)$ plane. The reflection extension 
used to define $(\mathcal{T},\mathcal{X})$ preserves these null trajectories 
continuously across the throat.

An important feature of the conformal diagram is that the throat is 
represented by a regular timelike surface rather than a spacetime singularity. 
As discussed in Sec.~\ref{sec3}, all curvature invariants remain finite 
throughout the spacetime and no additional singular boundaries arise in the 
conformal extension. The Penrose diagram, therefore, confirms that the wormhole 
geometry is globally regular and geodesically complete across the throat, 
providing a smooth bridge between two asymptotically flat regions.

\section{Quasinormal Modes}\label{sec7}
Quasinormal modes (QNMs) describe the characteristic damped oscillations of a 
perturbed compact object and are characterized by complex frequencies, whose 
real and imaginary parts represent the oscillation frequency and damping rate, 
respectively. Over the past few decades, QNMs have become an important tool 
for probing the stability and observational signatures of compact objects. 
Comprehensive reviews on QNMs can be found in Refs.~\cite{a42,a68,a69,a70,a71,
a72,a73,a74,a75,a76,a77,a78}. Here, we discuss the QNMs of the 5D EGB wormhole 
for two categories, viz., eikonal QNMs, which are based on the correspondence
of the photon-sphere dynamics with the perturbation of the related compact 
object, and scalar QNMs, where a physical scalar field is considered as a 
probe for the perturbation of the associated compact object.     

\subsection{Eikonal Quasinormal Modes}\label{sec7sub1}

The eikonal QNMs correspond to the large angular momentum 
limit ($l\gg1$), where the propagation of perturbations in the object is 
closely related to the properties of unstable circular null geodesics. In this 
regime, the real part of the QNM frequency is determined by the angular 
velocity of the photon-sphere, whereas the imaginary part is controlled by the 
instability of the orbit, quantified by the Lyapunov exponent. This 
correspondence establishes a direct link between the wave dynamics of 
spacetime and the behavior of photons moving along unstable circular 
trajectories~\cite{a66,a67}.

For the static and spherically symmetric wormhole metric~\eqref{eq21}, the 
radius of the unstable photon-sphere, $r_\text{ph}$, is obtained from the 
condition~\eqref{eq75}. The corresponding angular velocity of photons is 
given by
\begin{equation}
\Omega_c=\frac{\sqrt{f(r_\text{ph})}}{r_\text{ph}},
\end{equation}
while the Lyapunov exponent, which measures the instability timescale of the 
circular null orbit, is expressed as
\begin{equation}
\lambda_\text{p}=\sqrt{\frac{f(r_\text{ph})\left[2f(r_\text{ph})-r_\text{ph}^{2}f''(r_\text{ph})\right]}
{2r_\text{ph}^{2}}}.
\end{equation}
\begin{table}[!h]
\centering
\caption{Eikonal quasinormal frequencies $\omega$ for the fundamental mode 
($n=0$). The upper panel shows the variation with the Gauss-Bonnet coupling 
parameter $\alpha$ for fixed $\rho_0=1$, while the lower panel shows the 
variation with the string cloud parameter $\rho_0$ for fixed $\alpha=0.25$.}
\vspace{5pt}
\label{tab4}
\renewcommand{\arraystretch}{1.2}

\begin{tabular}{ccccc}
\toprule
$\alpha$ & $l=1$ & $l=2$ & $l=3$ & $l=4$ \\
\midrule
0.1 &
$0.397736-0.208822i$ &
$0.795472-0.208822i$ &
$1.19321-0.208822i$ &
$1.59094-0.208822i$ \\

0.2 &
$0.375381-0.183561i$ &
$0.750762-0.183561i$ &
$1.12614-0.183561i$ &
$1.50152-0.183561i$ \\

0.3 &
$0.357061-0.164461i$ &
$0.714121-0.164461i$ &
$1.07118-0.164461i$ &
$1.42824-0.164461i$ \\

0.4 &
$0.341628-0.149426i$ &
$0.683255-0.149426i$ &
$1.02488-0.149426i$ &
$1.36651-0.149426i$ \\
\bottomrule
\end{tabular}

\vspace{0.2cm}

\begin{tabular}{ccccc}
\toprule
$\rho_0$ & $l=1$ & $l=2$ & $l=3$ & $l=4$ \\
\midrule
0 &
$0.428373-0.247321i$ &
$0.856746-0.247321i$ &
$1.28512-0.247321i$ &
$1.71349-0.247321i$ \\

1 &
$0.365801-0.173400i$ &
$0.731602-0.173400i$ &
$1.097404-0.173400i$ &
$1.46320-0.173400i$ \\

2 &
$0.297338-0.112519i$ &
$0.594676-0.112519i$ &
$0.892014-0.112519i$ &
$1.18935-0.112519i$ \\

2.5 &
$0.261966-0.091338i$ &
$0.523931-0.091338i$ &
$0.785897-0.091338i$ &
$1.04786-0.091338i$ \\
\bottomrule
\end{tabular}
\end{table}
According to the geodesic correspondence established in the eikonal limit, the QNM frequencies at leading order are expressed as \cite{a66}
\begin{equation}
\omega_{ln}
=l \Omega_c-i\left(n+\frac{1}{2}\right)\lambda_\text{p},
\label{eq87}
\end{equation}
where $l$ denotes the angular momentum number and $n=0,1,2,\ldots$ is the
overtone number. The real and imaginary parts of the QNM frequency correspond
to the oscillation frequency and the damping rate of the perturbation,
respectively. To examine the effects of the Gauss-Bonnet coupling parameter
$\alpha$ and the string cloud parameter $\rho_0$, we numerically determine the
photon-sphere radius and evaluate the corresponding eikonal QNM frequencies
for different parameter values. The fundamental ($n=0$) eikonal QNM
frequencies are listed in Table~\ref{tab4}. The upper panel shows the
variation with the Gauss-Bonnet coupling $\alpha$ for a fixed string cloud
density $\rho_0=1$, whereas the lower panel illustrates the effect of the
string cloud parameter $\rho_0$ for a fixed coupling $\alpha=0.25$.

For fixed model parameters, the real part of the quasinormal frequency
increases linearly with the angular momentum number $l$, while the imaginary
part remains unchanged. This behaviour follows directly from the eikonal
relation~\eqref{eq87}, since both the angular velocity $\Omega_c$ and the
Lyapunov exponent $\lambda_\text{p}$ are determined solely by the properties
of the unstable photon orbit and are therefore independent of $l$.

The upper panel of Table~\ref{tab4} shows that increasing the Gauss-Bonnet
coupling parameter $\alpha$ leads to a decrease in both the real and imaginary
parts of the QNM frequency. A smaller real part corresponds to a lower
oscillation frequency, whereas a reduced magnitude of the imaginary part
indicates that the perturbations decay more slowly. Therefore, stronger
Gauss-Bonnet corrections give rise to longer-lived QNM oscillations. A similar
behaviour is observed in the lower panel when the string cloud parameter
$\rho_0$ is increased. Both the oscillation frequency and the damping rate
decrease monotonically with increasing $\rho_0$, suggesting that a denser
string cloud weakens the instability of the photon-sphere and consequently
extends the lifetime of the QNM ringing. Overall, both the Gauss-Bonnet
curvature corrections and the string cloud matter distribution suppress the
oscillation frequency while simultaneously reducing the damping rate of the
eikonal QNMs.

\subsection{Scalar Quasinormal Modes}\label{sec7sub2}

In this subsection, we study the dynamical stability of the traversable 
wormhole against massless scalar field perturbations through its QNM spectrum.
For this analysis, we adopt the test-field approximation, in which the 
scalar field evolves on the fixed wormhole background without modifying the 
spacetime geometry. Consequently, the backreaction of the perturbing field, as 
well as quantum effects and the influence of distant matter fields, are 
neglected~\cite{a55,a79}. Depending on the physical scenario, one may consider 
scalar, electromagnetic, gravitational or fermionic (Dirac) 
perturbations~\cite{a80,a81,a82}. In the present work, we restrict our 
attention to massless scalar perturbations.

\begin{figure}[!h]
\centering
\includegraphics[scale=0.6]{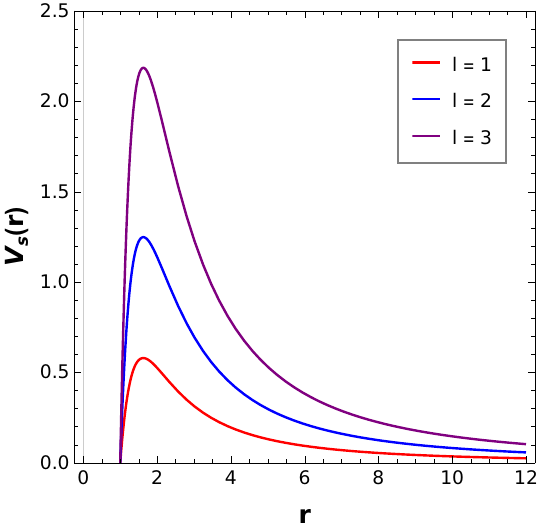}\hspace{0.5cm}
\includegraphics[scale=0.6]{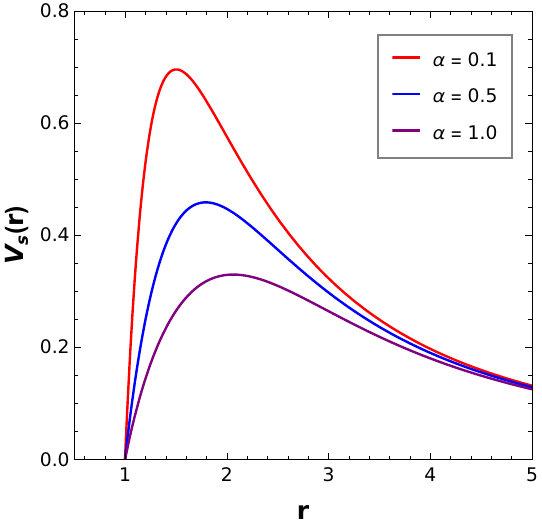}\hspace{0.5cm}
\includegraphics[scale=0.6]{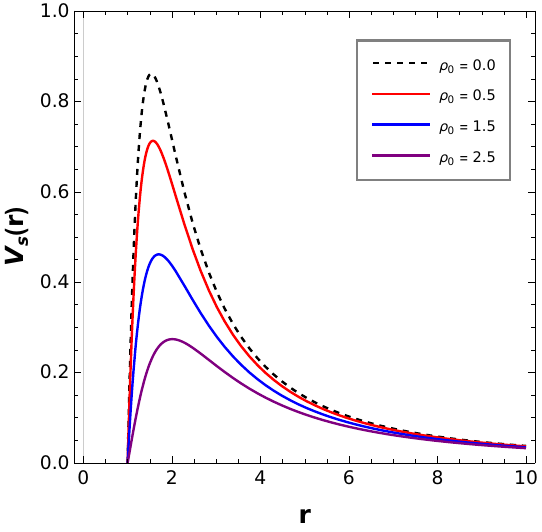}
\vspace{-0.2cm}
\caption{Effective scalar perturbation potential $V_s(r)$ for the 5D EGB 
wormhole with $r_0=1$. The left panel illustrates the dependence on the 
multipole number $l$ for fixed $(\alpha,\rho_0)=(0.25,1)$. The middle panel 
shows the variation with the Gauss-Bonnet coupling $\alpha$ for fixed 
$(\rho_0,l)=(1,1)$, while the right panel presents the effect of the string 
cloud parameter $\rho_0$ for fixed $(\alpha,l)=(0.25,1)$.}
\label{fig11}
\end{figure}

To analyze the evolution of the perturbation, we introduce the tortoise 
coordinate, which recasts the radial part of the wave equation into a 
Schr\"odinger-like equation with an effective potential. Accordingly, we 
consider a massless scalar field $\zeta$ propagating on the wormhole 
background while neglecting its backreaction on the spacetime geometry. The 
field satisfies the massless Klein-Gordon equation~\cite{a44}
\begin{equation}
\Box \zeta = \frac{1}{\sqrt{-g}} \partial_\mu \left( \sqrt{-g}\, g^{\mu\nu} \partial_\nu \zeta \right) = 0.
\label{eq88}
\end{equation}
Expanding the scalar field in scalar spherical harmonics on the unit 
three-sphere ($S^3$),
\begin{equation}
\zeta(t,r,\xi,\theta,\phi)=e^{-i\omega t}\frac{\psi(r)}{r^{3/2}}Y^\ell_m(\xi,\theta,\phi),
\label{eq89}
\end{equation}
the Klein-Gordon equation reduces to the Schr\"{o}dinger-like wave equation as
\begin{equation}
\frac{d^2\psi(r_*)}{dr_*^2}+\left[\omega^2-V_s(r_*)\right]\psi(r_*)=0,
\label{eq90}
\end{equation}
where the tortoise coordinate $r_*$ is defined by
\begin{equation}
\frac{dr_*}{dr}=\pm \frac{1}{f(r)},
\label{eq91}
\end{equation}
and the effective potential is
\begin{equation}
V_s(r)=f(r)\left[\frac{l(l+2)}{r^2}+\frac{3f(r)}{4r^2}+\frac{3f'(r)}{2r}\right],
\label{eq92}
\end{equation}
with $l$ denoting the multipole number. For the present wormhole solution, the 
effective potential forms a single barrier in the vicinity of the throat and 
vanishes asymptotically in both directions. Consequently, the QNMs satisfy 
the boundary conditions~\cite{a83}:
\begin{equation}
\psi(r_*)\sim
\begin{cases}
e^{-i\omega r_*}, & r_*\rightarrow+\infty,\\
e^{+i\omega r_*}, & r_*\rightarrow-\infty,
\end{cases}
\label{eq93}
\end{equation}
which correspond to purely outgoing waves at both asymptotic infinities. 
Fig.~\ref{fig11} presents the effective scalar potential $V_s(r)$ for the 5D 
EGB wormhole with the throat radius fixed at $r_0=1$. In each case, the 
potential forms a single positive barrier outside the throat and decreases 
monotonically toward zero as the radial coordinate approaches either 
asymptotically flat regions, satisfying the conditions required for the 
application of the WKB approximation. The left panel illustrates the 
dependence of the potential on the multipole number $l$ for fixed 
$(\alpha,\rho_0)=(0.25,1)$. As $l$ increases, the height of the potential 
barrier rises significantly, whereas the position of its maximum changes only 
marginally, indicating that higher multipole modes encounter a stronger 
scattering barrier. The middle panel shows the effect of the Gauss-Bonnet 
coupling parameter $\alpha$ for fixed $(\rho_0,l)=(1,1)$. Increasing $\alpha$ 
lowers the height of the potential barrier and shifts its peak slightly away 
from the throat, implying that the Gauss-Bonnet correction weakens the 
effective barrier experienced by the scalar field. The right panel depicts 
the influence of the string cloud parameter $\rho_0$ for fixed 
$(\alpha,l)=(0.25,1)$. The vacuum configuration ($\rho_0=0$) exhibits the 
highest potential barrier, while increasing $\rho_0$ progressively reduces 
its height. This behaviour suggests that the presence of the string cloud 
weakens the effective scattering potential near the wormhole throat.

\begin{figure}[!h]
\centering
\includegraphics[scale=0.6]{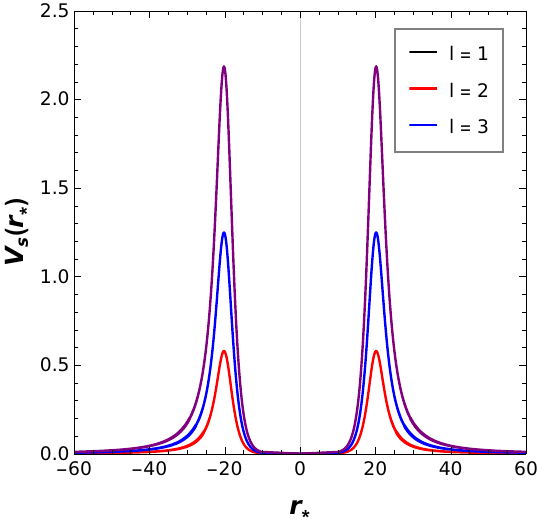}\hspace{0.5cm}
\includegraphics[scale=0.6]{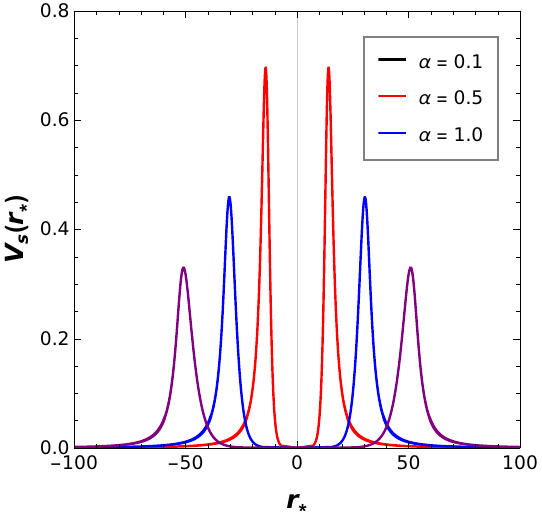}\hspace{0.5cm}
\includegraphics[scale=0.58]{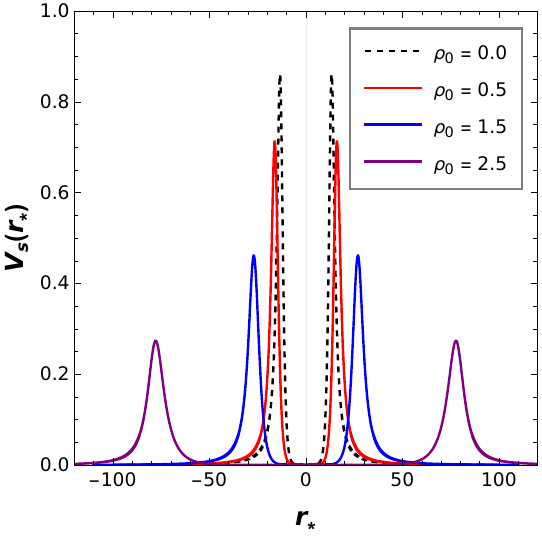}
\vspace{-0.2cm}
\caption{Effective scalar perturbation potential $V_s(r_*)$ for the 5D EGB 
wormhole with $r_0=1$. The left panel illustrates the dependence on the 
multipole number $l$ for fixed $(\alpha,\rho_0)=(0.25,1)$. The middle panel 
shows the variation with the Gauss-Bonnet coupling $\alpha$ for fixed 
$(\rho_0,l)=(1,1)$, while the right panel presents the effect of the string 
cloud parameter $\rho_0$ for fixed $(\alpha,l)=(0.25,1)$.}
\label{fig12}
\end{figure}

Fig.~\ref{fig12} shows the effective scalar potential $V_s(r_*)$ in terms of 
the tortoise coordinate $r_*$. Because of the two asymptotically flat regions 
connected by the wormhole throat, the potential develops a symmetric 
double-barrier profile, with identical peaks located on both sides of the 
throat and approaches zero as $r_*\rightarrow\pm\infty$. The left panel 
demonstrates that the barrier height increases noticeably with the multipole 
number $l$, while the separation between the two peaks remains nearly 
unchanged. In contrast, the middle and right panels reveal that increasing 
either the Gauss-Bonnet coupling $\alpha$ or the string cloud parameter 
$\rho_0$ lowers the barrier height and shifts the peaks farther away from the 
throat in the tortoise coordinate. These results indicate that larger values 
of $l$ strengthen the effective scattering potential, whereas both the 
Gauss-Bonnet corrections and the string cloud matter distribution weaken it. 
Since the QNM spectrum is governed by the shape of the effective potential, 
variations in $\alpha$ and $\rho_0$ are expected to produce noticeable changes 
in the oscillation frequencies and damping rates of the scalar perturbations.

Since the effective potential $V_s(r)$ consists of a single smooth barrier, 
the WKB approximation provides a suitable semianalytical approach for 
determining the corresponding QNM frequencies. This method is well known to 
produce accurate results for single-barrier potentials, particularly in the 
regime where the multipole number exceeds the overtone number ($l>n$). The 
WKB formalism was first introduced by Schutz and Will~\cite{a83a} and later 
extended to higher orders by Iyer, Will, and Konoplya~\cite{a84,a85}. Its 
accuracy can be further enhanced by employing Pad\'{e} approximations, which 
significantly improve the convergence of the WKB series~\cite{a86}.

In the present work, we compute the QNM frequencies of massless scalar 
perturbations using the sixth-order Pad\'{e}-improved WKB method. To assess 
the reliability of the obtained frequencies, we also estimate the numerical 
uncertainty associated with the WKB approximation using the standard 
expression adopted in Refs.~\cite{a85,a87,a88} as given by
\begin{equation}
\Delta_6=\frac{\left|\text{WKB}_7-\text{WKB}_5\right|}{2},
\label{eq94}
\end{equation}
where $\text{WKB}_7$ and $\text{WKB}_5$ denote the seventh and fifth-order WKB 
results, respectively. This quantity provides an estimate of the truncation 
error associated with the finite-order WKB expansion, with smaller values of 
$\Delta_6$ indicating better convergence of the method. To quantify the 
agreement between the sixth-order Pad\'e WKB results and the eikonal 
approximation, we define the relative difference between the corresponding 
complex QNM frequencies as
\begin{equation}
\Delta_{\text{rel}}=
\left|\frac{\text{QNM}_{\text{Eik}}-\text{QNM}_{\text{WKB}}}
{\text{QNM}_{\text{Eik}}}\right|,
\label{eq95}
\end{equation}
where $\text{QNM}_{\text{WKB}}$ denote the sixth-order QNMs obtained from the 
Pad\'{e} averaged WKB method and $\text{QNM}_{\text{Eik}}$ denote the eikonal 
QNMs, respectively. Smaller values of $\Delta_{\text{rel}}$ indicate better 
agreement between the two approaches.

Table~\ref{tab5} lists the sixth-order Pad\'{e} averaged WKB quasinormal 
frequencies together with the corresponding eikonal frequencies, the numerical 
uncertainty $\Delta_6$, and the relative difference $\Delta_{\text{rel}}$ for 
different multipole numbers $l$. The parameters are fixed at $r_0=1$, 
$\rho_0=1$ and $\alpha=0.25$. As the multipole number increases, the real part 
of the quasinormal frequency increases almost linearly, while the imaginary 
part remains nearly unchanged. This behaviour indicates that higher multipole 
modes oscillate at higher frequencies but decay at nearly the same rate.
The numerical uncertainty $\Delta_6$ decreases rapidly with increasing $l$. 
For the lowest multipole ($l=1$), the uncertainty is relatively large, 
reflecting the reduced accuracy of the WKB approximation in the low-$l$ 
regime. In contrast, for $l\ge2$, the uncertainty decreases by several orders 
of magnitude, demonstrating the excellent convergence of the higher-order WKB 
expansion. This behaviour is consistent with the well-established validity of 
the WKB method when the multipole number exceeds the overtone number ($l>n$).
The relative difference $\Delta_{\text{rel}}$ between the WKB and eikonal 
frequencies also decreases monotonically with increasing $l$, indicating 
progressively better agreement between the two approaches. This trend is 
expected because the eikonal approximation becomes increasingly accurate in 
the large-$l$ limit, where the QNM spectrum is governed by the properties of 
the unstable photon-sphere. Consequently, the sixth-order WKB frequencies 
converge toward the corresponding eikonal values as the multipole number 
increases.

\begin{table}[ht]
\centering
\caption{Sixth-order WKB quasinormal frequencies ($\text{QNM}_{\text{WKB}}$), the 
corresponding eikonal frequencies ($\text{QNM}_{\text{Eik}}$), the numerical 
uncertainty $\Delta_6$ and the relative difference $\Delta_{\text{rel}}$ for 
different multipole numbers $l$. The parameters are fixed at $r_0=1$, 
$\rho_0=1$ and $\alpha=0.25$.}
\vspace{5pt}
\label{tab5}
\begin{tabular}{ccccc}
\hline\hline
$l$ & $\text{QNM}_{\text{WKB}}$ & $\text{QNM}_{\text{Eik}}$ &$\Delta_6$  & $\Delta_{\text{rel}}$ \\[1pt]
\hline
1 & $0.729249 - 0.177416i$ & $0.365801 - 0.173400i$ & $0.44491$ & $0.89786$ \\
2 & $1.095276 - 0.174979i$ & $0.731602 - 0.173400i$ & $0.04536$ & $0.483696$ \\
3 & $1.461743 - 0.174238i$ & $1.097404 - 0.173400i$ & $0.65130 \times 10^{-2}$ & $0.327926$ \\
4 & $1.827908 - 0.173932i$ & $1.463205 - 0.173400i$ & $0.90460 \times 10^{-4}$ & $0.247518$ \\
5 & $2.193909 - 0.173772i$ & $1.829006 - 0.173400i$ & $0.69355 \times 10^{-5}$ & $0.198619$ \\
6 & $2.559855 - 0.173676i$ & $2.194807 - 0.173400i$ & $0.77908 \times 10^{-4}$ & $0.165807$ \\
\hline\hline
\end{tabular}
\end{table}

\begin{figure}[!h]
\centering
\includegraphics[scale=0.65]{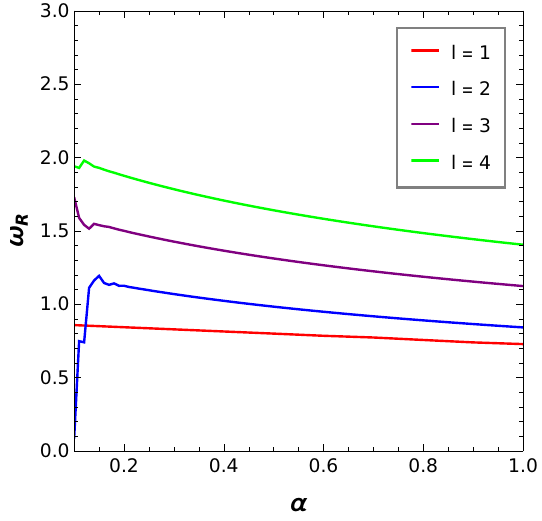}\hspace{1.0cm}
\includegraphics[scale=0.66]{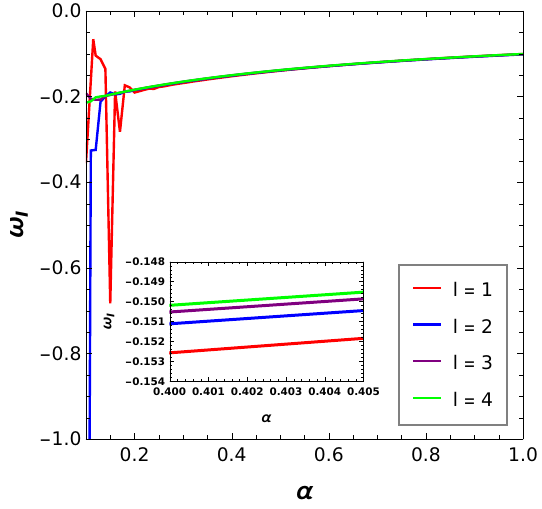}
\vspace{-0.2cm}
\caption{Variation of the real (left) and imaginary (right) parts of the 
fundamental QNM frequencies with the Gauss-Bonnet coupling parameter 
$\alpha$ for different multipole numbers $l$, with $\rho_0=1$ and $r_0=1$.}
\label{fig13}
\end{figure}

Fig.~\ref{fig13} shows the variation of the QNM frequencies with the 
Gauss-Bonnet coupling parameter $\alpha$ for different multipole numbers $l$. 
For all multipoles considered, the real part of the QNM frequency decreases 
gradually as $\alpha$ increases, indicating that the oscillation frequency of 
the scalar perturbations is reduced by stronger Gauss-Bonnet corrections. At 
the same time, the magnitude of the imaginary part also decreases, implying 
that the perturbations decay more slowly and the corresponding QNM 
oscillations persist for longer times. This dependence is more pronounced for 
the lower multipole modes, while the higher multipoles vary more smoothly with 
$\alpha$. Small irregularities are observed in the WKB frequencies for very 
small values of $\alpha$, particularly for the lowest multipoles. These 
features are most likely associated with the reduced accuracy of the WKB 
approximation in this parameter regime. Overall, increasing the Gauss-Bonnet 
coupling lowers both the oscillation frequency and the damping rate of the 
scalar QNMs.

\begin{figure}[!h]
\centering
\includegraphics[scale=0.65]{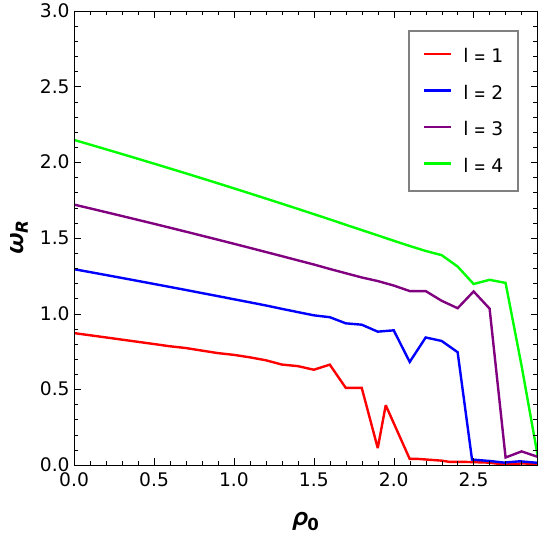}\hspace{1.0cm}
\includegraphics[scale=0.68]{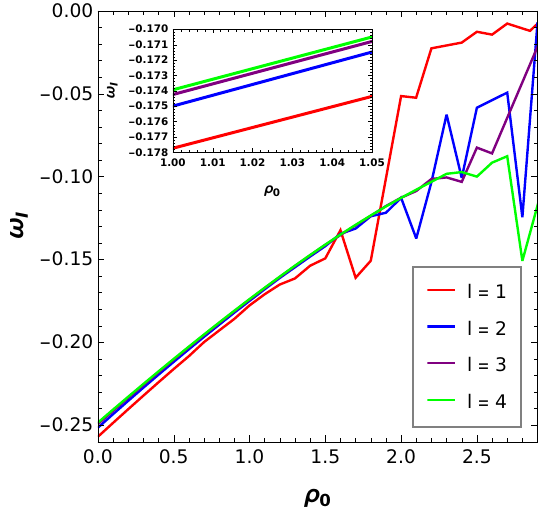}
\vspace{-0.2cm}
\caption{Variation of the real (left) and imaginary (right) parts of the 
fundamental QNM frequencies with the string cloud parameter $\rho_0$ for 
different multipole numbers $l$, with $\alpha=0.25$ and $r_0=1$.}
\label{fig14}
\end{figure}

Fig.~\ref{fig14} presents the dependence of the QNM frequencies on the string 
cloud parameter $\rho_0$. As $\rho_0$ increases, the real part of the 
frequency decreases monotonically for all multipole numbers, indicating a 
progressive reduction in the oscillation frequency of the perturbations. 
For $\rho_0 \gtrsim 1.5$, this decrease becomes particularly significant for 
the lower multipoles, with the real part approaching zero. The imaginary part 
also decreases smoothly over a wide range of $\rho_0$, indicating a slower 
decay of the perturbations. However, for larger values of $\rho_0$, noticeable 
irregularities appear in the WKB frequencies, especially for the lower 
multipoles. Since the effective potential $V_s(r)$ in the physical exterior 
region continues to exhibit a single dominant barrier, these irregular 
features are unlikely to arise from a change in the qualitative structure of 
the potential. Instead, they may reflect the reduced reliability of the WKB 
approximation in the regime when the potential barrier becomes relatively low 
and broad.

\subsection{Evolution of Scalar Perturbation and Time-domain Analysis}
\label{sec7sub3}

\begin{figure}[!h]
\centering
\includegraphics[scale=0.35]{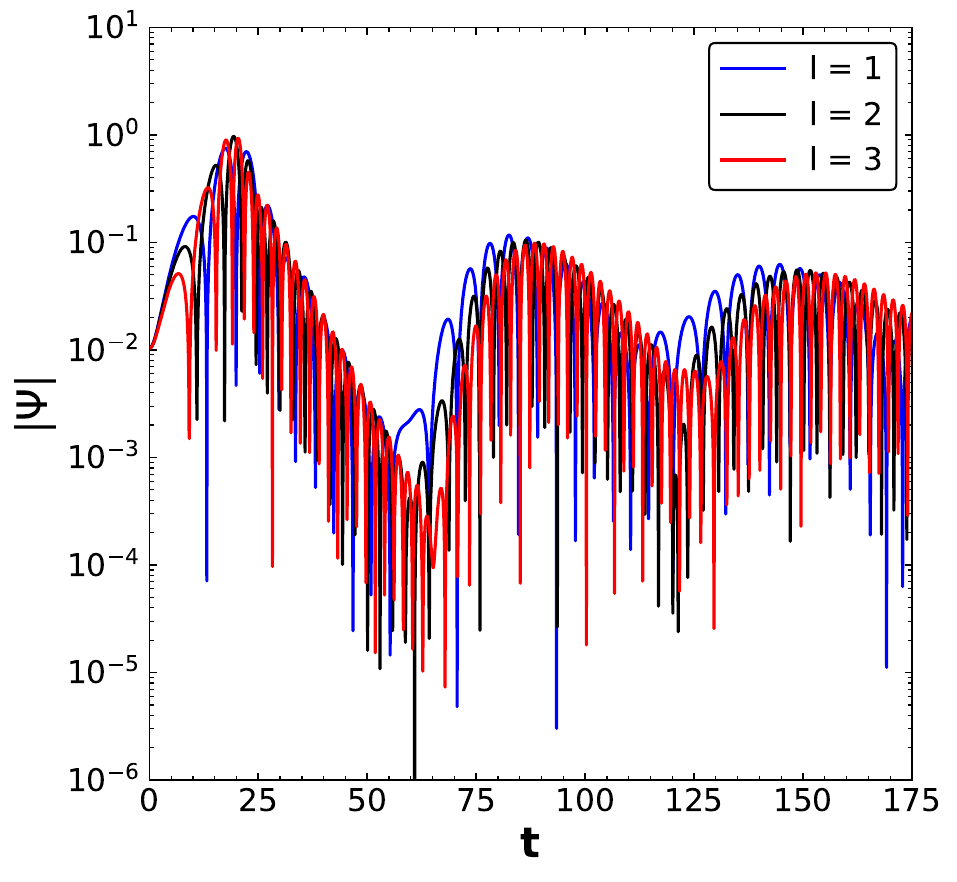}\hspace{1.0cm}
\includegraphics[scale=0.35]{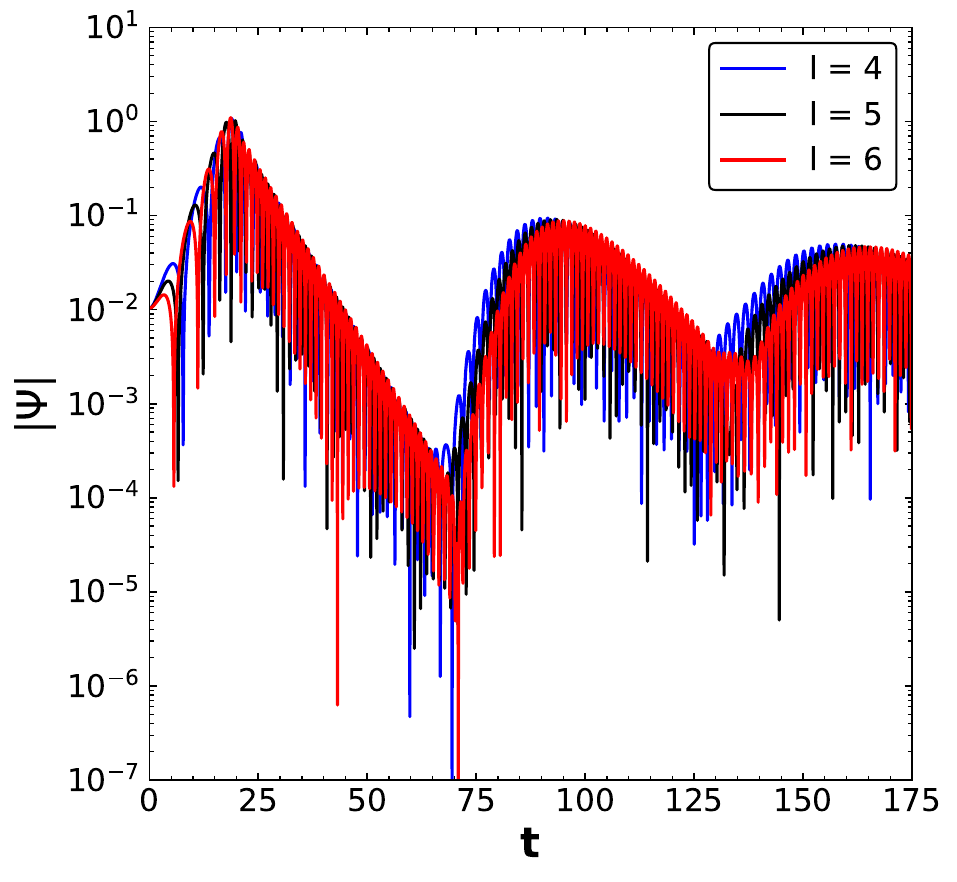}
\vspace{-0.2cm}
\caption{Time-domain evolution of scalar perturbations for multipole numbers 
$l=1$ to $6$ with $\alpha=0.25$, $\rho_0=1$, and $r_0=1$.}
\label{fig15}
\end{figure}

To further examine the dynamical stability of the wormhole and to validate 
the QNM frequencies obtained from the WKB approximation, we perform a 
time-domain evolution analysis of massless scalar perturbations. The evolution 
study is carried out using the time-domain integration method developed in 
Refs.~\cite{a89,a90}. Discretizing the wave function as 
$\Psi(r_*,t)=\Psi(i\Delta r_*,j\Delta t)\equiv\Psi_{i,j}$ and the effective 
potential as $V(r(r_*))\equiv V_i$, the wave equation~\eqref{eq88} can be
written in the form:
\begin{equation}
\frac{\Psi_{i+1,j} - 2\Psi_{i,j} + \Psi_{i-1,j}}{\Delta r_*^2}-\frac{\Psi_{i,j+1} - 2\Psi_{i,j} + \Psi_{i,j-1}}{\Delta t^2} - V_i \Psi_{i,j} = 0.
\label{eq96}
\end{equation}
We consider initial wave as a Gaussian wave packet of the form
$\Psi(r_*, t) = \exp\left[-(r_* - k)^2/2\sigma^2\right]$ and the 
condition $\left.\partial \Psi(r_*, t)/\partial t\right|_{t<0} = 0$, 
where $k$ and  $\sigma$ are the median and width of the initial wave-packet. 
Using these initial conditions, it is possible to express the time evolution 
of the scalar field as
\begin{equation}
\Psi_{i,j+1} = -\Psi_{i,j-1}
+ \left(\frac{\Delta t}{\Delta r_*}\right)^2
\left(\Psi_{i+1,j} + \Psi_{i-1,j}\right)
+ \left(2 - 2\left(\frac{\Delta t}{\Delta r_*}\right)^2 - V_i \Delta t^2 \right)\Psi_{i,j}. \label{eq97}
\end{equation}
To ensure the numerical stability, the time step is chosen to satisfy the 
von Neumann stability condition, $\Delta t/\Delta r_*<1$, throughout the 
analysis. Fig.~\ref{fig15} presents the time-domain evolution of massless scalar 
perturbations for multipole numbers $l=1$ to $6$, with the wormhole parameters 
fixed at $\alpha=0.25$, $\rho_0=1$, and $r_0=1$. After an initial transient 
stage, the perturbations enter the characteristic QNM ringing phase, where 
the signal is dominated by damped oscillations. As the multipole number 
increases, the oscillation frequency becomes higher, resulting in a greater 
number of oscillation cycles over the same time interval. This behaviour is 
in good agreement with the WKB results, which predict an increase in the real 
part of the quasinormal frequencies with increasing $l$.
At later times, the waveforms display a sequence of delayed oscillatory pulses 
with progressively decreasing amplitudes, a characteristic signature of 
echo-like behaviour. These echoes originate from repeated partial reflections 
of the scalar waves between the two symmetric potential barriers of the 
effective potential $V_s(r_*)$, producing successive delayed signals after 
the primary QNM ringdown~\cite{a45a}. Throughout the evolution, the amplitude 
of the perturbations decreases monotonically, and no exponentially growing 
modes are observed. This indicates that the 5D EGB wormhole remains 
dynamically stable against massless scalar perturbations for the range of 
parameters considered.

To extract the QNM frequencies from the numerical time evolution profiles, 
the primary ringdown portion of each waveform was fitted using the 
Levenberg-Marquardt algorithm \cite{a35a,a91,a92,a93} with the damped 
sinusoidal function $\Psi(t)=A e^{-\omega_I t}\cos(\omega_R t+\phi)$.
\begin{figure}[!h]
\centerline{
\includegraphics[scale=0.33]{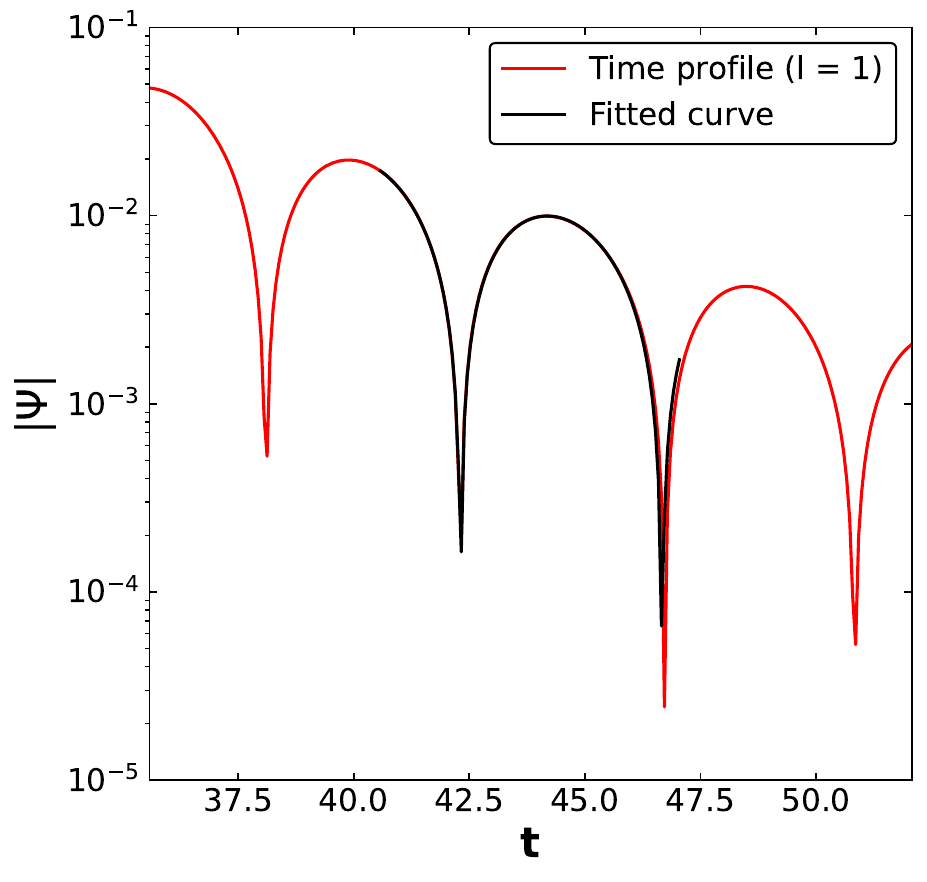}\hspace{0.5cm}
\includegraphics[scale=0.33]{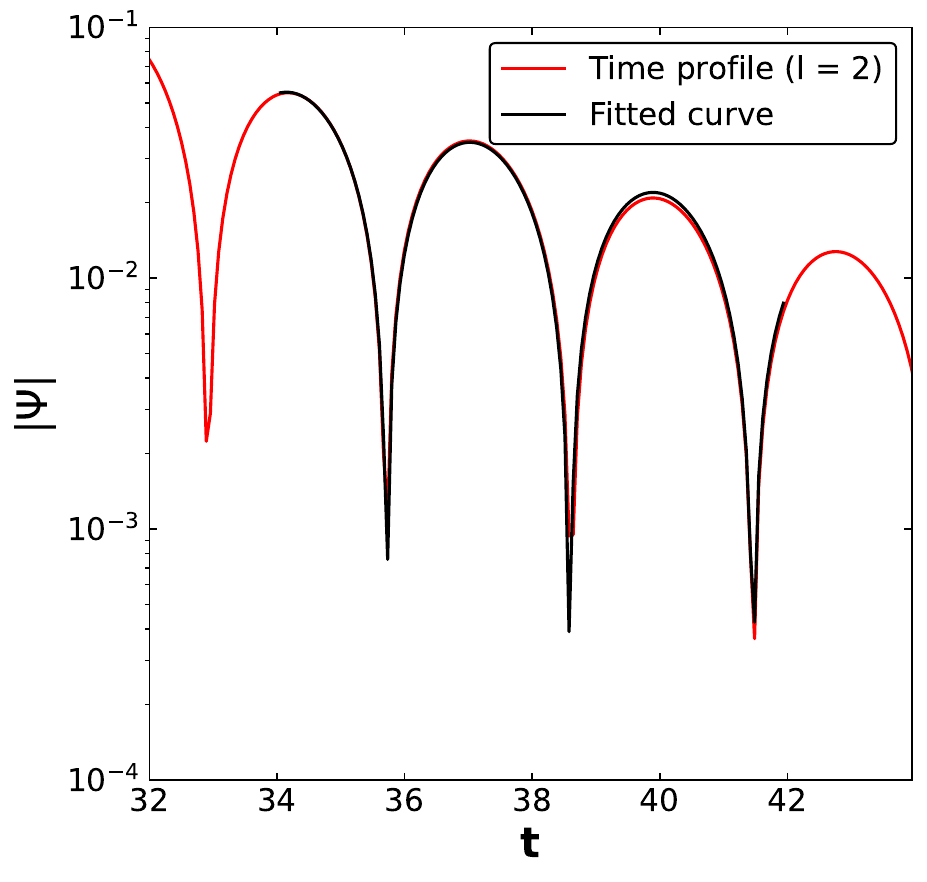}\hspace{0.5cm}
\includegraphics[scale=0.33]{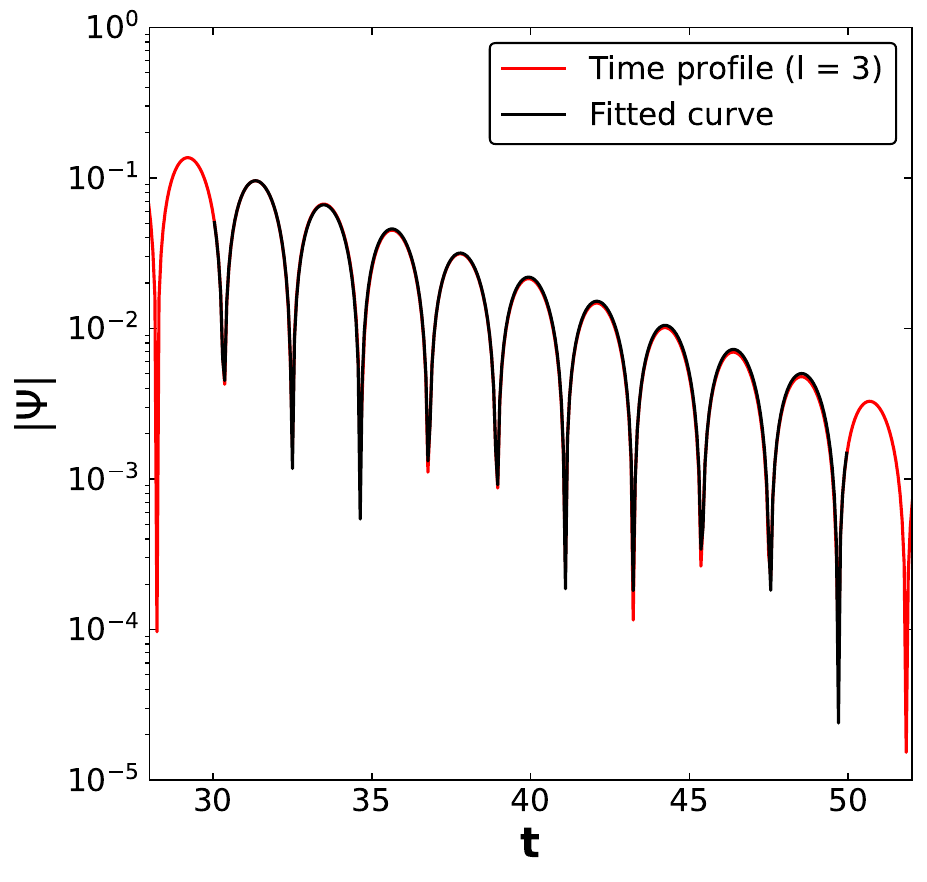}\hspace{0.5cm}}
\centerline{
\includegraphics[scale=0.33]{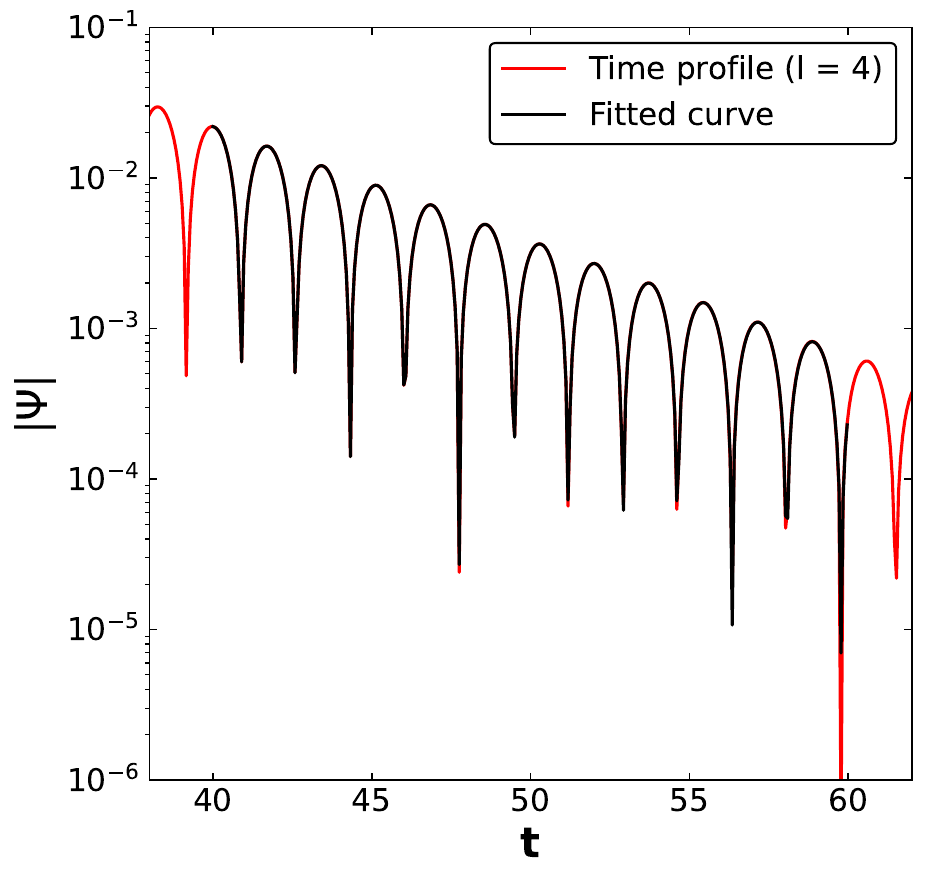}\hspace{0.5cm}
\includegraphics[scale=0.33]{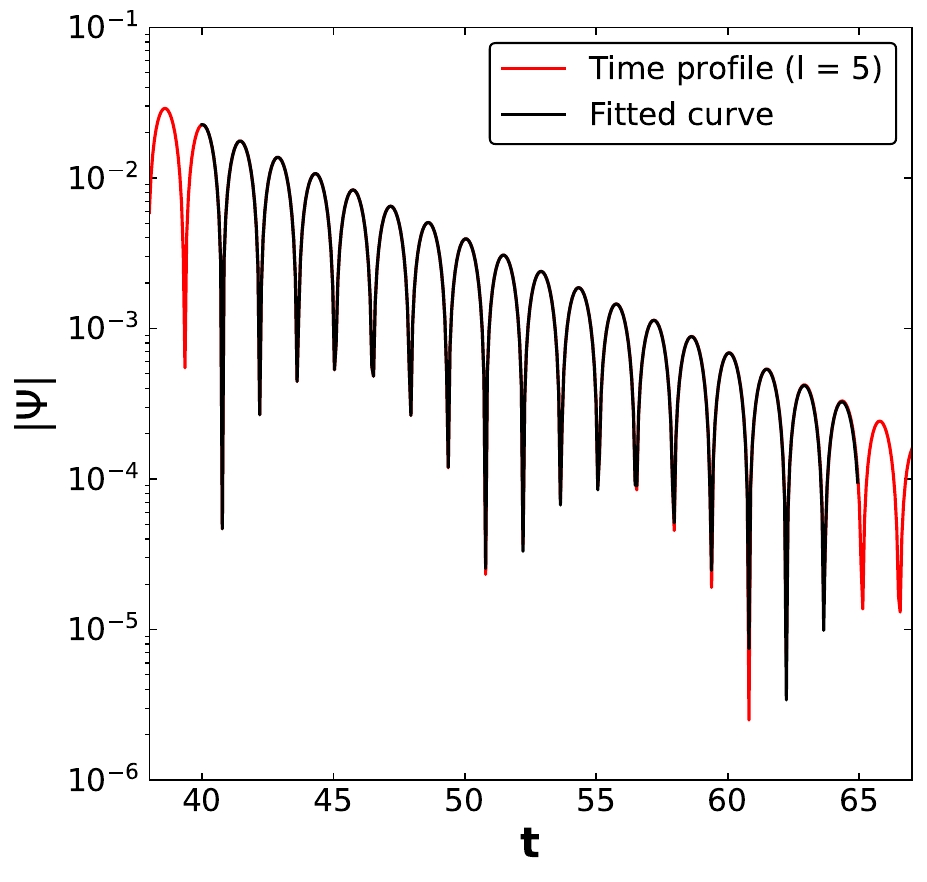}\hspace{0.5cm}
\includegraphics[scale=0.33]{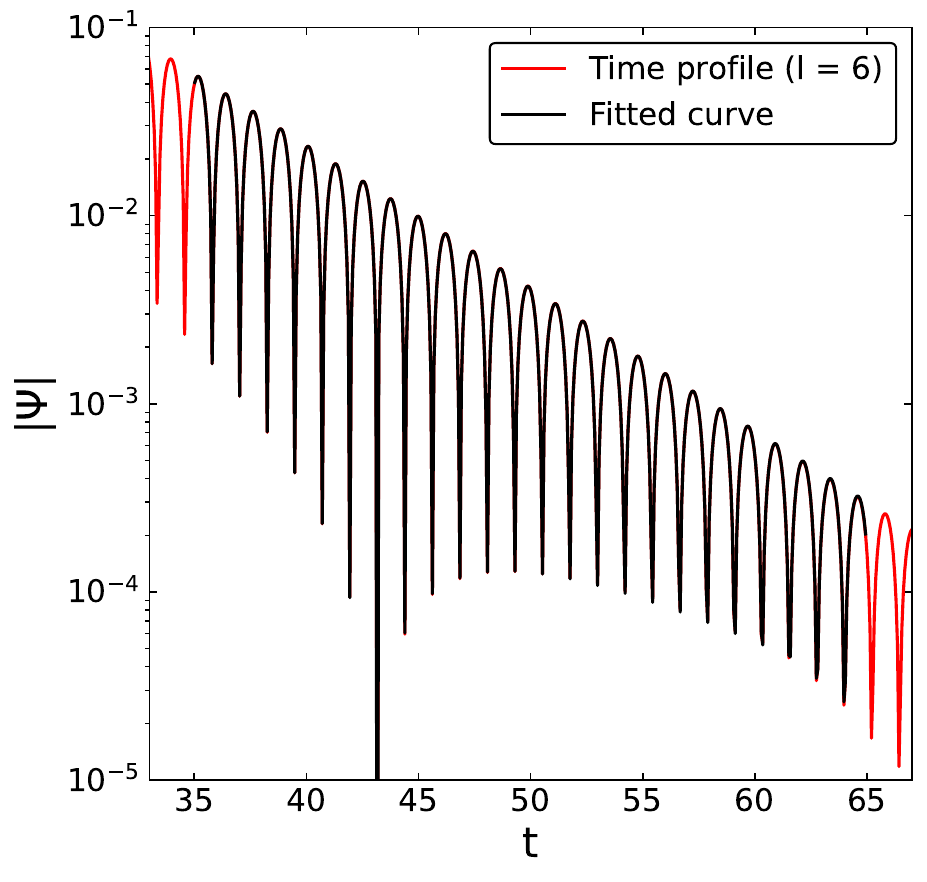}\hspace{0.5cm}}
\vspace{-0.2cm}
\caption{Fitting of the time-domain profiles to estimate the QNMs for 
multipole numbers $l=1$ to $6$. The fitting is performed in the QNM ringing 
regime for the wormhole parameters $\alpha=0.25$, $\rho_0=1$, and $r_0=1$.}
\label{fig16}
\end{figure}
The fitting was carried out over the time interval of $20 \leq t \leq 65$, 
which corresponds to the QNM ringing phase following the initial transient 
and preceding the onset of the late-time echo-like oscillations. 
Fig.~\ref{fig16} compares the numerical time-domain waveforms with the 
corresponding fitted profiles for multipole numbers $l=1$ to $6$. The fitted 
curves closely reproduce the numerical evolution throughout the chosen 
fitting interval, demonstrating that the dominant ringdown signal is well 
described by the fundamental QNM. The agreement between the numerical data 
and the fitted profiles also suggests that the contributions from higher 
overtones and the late-time echo-like signals are negligible within the 
selected fitting window. To compare the time-domain results 
with those obtained from the sixth-order Pad\'{e}-improved WKB 
approximation, we define the difference between the corresponding complex 
QNM frequencies as
\begin{equation}
\Delta_{\text{QNM}} = \frac{\left| \text{QNM}_{\text{WKB}} - \text{QNM}_{\text{Time-domain}} \right|}{2}. \label{eq98}
\end{equation}
Smaller values of $\Delta_{\text{QNM}}$ indicate better agreement 
between the two approaches. The coefficient of determination is computed as 
\begin{equation}
R^2 = 1 - \frac{\sum \left( \Psi_{\text{Time\text{-}Profile}} - \Psi_{\text{Fit}} \right)^2}{\sum \left( \Psi_{\text{Time\text{-}Profile}} - \bar{\Psi}_{\text{Time\text{-}Profile}} \right)^2}, \label{eq99}
\end{equation}
where $\Psi_{\text{Time-Profile}}$ is the amplitude of the original 
time-domain waveform, $\Psi_{\text{Fit}}$ is the amplitude of the fitted 
waveform and $\bar{\Psi}_{\text{Time-Profile}}$ denotes the mean amplitude of 
the original time-domain waveform.
The QNM frequencies obtained from the time-domain analysis are summarized in 
Table~\ref{tab6}, together with the corresponding sixth-order Pad\'{e} 
averaged WKB results. Overall, the two independent approaches show good 
agreement for all multipole numbers considered. The difference between the 
two sets of frequencies, quantified by $\Delta_{\text{QNM}}$, decreases 
steadily from $l=1$ to $l=4$, reaching a minimum value of $6.65\times10^{-4}$, 
before exhibiting a slight increase for $l=5$ and $l=6$. Even so, the 
differences remain at the level of $10^{-3}$ - $10^{-2}$, indicating a good 
consistency between the frequency-domain WKB method and the time-domain 
evolution. The quality of the time-domain fitting is further reflected in 
the coefficient of determination, which satisfies $R^2>0.999$ for all 
multipole modes and becomes even closer to unity as $l$ increases. This 
demonstrates that the damped sinusoidal model accurately reproduces the 
numerical ringdown signal over the chosen fitting interval and confirms that 
the extracted frequencies are robust.
\begin{table}[htbp]
\caption{Comparison between the QNM frequencies obtained from the sixth-order 
WKB method and those extracted from the time-domain fitting for 
$\alpha=0.25$, $\rho_0=1$, and $r_0=1$.}
\vspace{5pt}
\begin{tabular}{c c c c c}
\hline\hline
Multipole number & 6th-order WKB QNMs & Time-Domain QNMs & $R^2$ & $\Delta_{\text{QNM}}$ \\
\hline
$l=1$ & $0.729249 - 0.177416i$ & $0.72065 - 0.161233i$ & $0.999825$ & $0.91642 \times 10^{-2}$\\
$l=2$ & $1.095276 - 0.174979i$ & $1.09375 - 0.160112i$ & $0.999454$ & $0.74723 \times 10^{-2}$\\
$l=3$ & $1.461743 - 0.174238i$ & $1.46161 - 0.171479i$ & $0.999821$ & $0.13806 \times 10^{-2}$\\
$l=4$ & $1.827908 - 0.173932i$ & $1.82920 - 0.174241i$ & $0.999999$ & $0.66549 \times 10^{-3}$\\
$l=5$ & $2.193909 - 0.173772i$ & $2.19606 - 0.174310i$ & $0.999999$ & $0.11092 \times 10^{-2}$\\
$l=6$ & $2.559855 - 0.173676i$ & $2.56322 - 0.174595i$ & $0.999999$ & $0.17426 \times 10^{-2}$\\
\hline\hline
\end{tabular}
\label{tab6}
\end{table}

\section{Summary and Conclusion}\label{sec8}

In this work, we have constructed an exact static and spherically symmetric 
traversable wormhole solution in 5D EGB gravity supported by a Letelier cloud 
of strings. By imposing the flare-out condition together with the positivity 
of the string-cloud density, we identified the physically admissible region of 
the parameter space. The resulting spacetime was shown to be asymptotically 
flat, while the analysis of the Ricci scalar, Ricci tensor squared and 
Kretschmann scalar confirmed that the geometry remains regular throughout the 
physical domain, including at the wormhole throat.

To examine the physical viability of the solution, we recast the field 
equations into an effective Einstein form and investigated the corresponding 
effective energy conditions. We found that the effective null, weak, strong 
and dominant energy conditions are satisfied or saturated at the throat, 
indicating that the higher-curvature Gauss-Bonnet contributions effectively 
support the wormhole geometry while the underlying string-cloud matter 
remains physically reasonable. In addition, the generalized TOV equation 
demonstrated that the wormhole is in mechanical equilibrium through the 
balance between the hydrostatic and anisotropic forces.

The traversability of the wormhole was studied through the proper radial 
distance, embedding diagrams, traversal time, proper acceleration, and tidal 
accelerations. The proper radial distance and embedding analysis revealed a 
smooth connection between the two asymptotically flat regions without any 
geometrical pathology. Although the coordinate traversal time diverges in the 
adopted static coordinate system, the proper traversal time experienced by a 
traveler remains finite. Furthermore, a freely falling traveler experiences 
no proper acceleration, while the tidal accelerations remain finite, 
satisfying the essential Morris-Thorne requirements for traversability.

We also investigated the optical properties of the wormhole by analyzing null 
geodesics, the unstable photon-sphere, and the corresponding shadow. Both the 
Gauss-Bonnet coupling and the string-cloud parameter were found to influence 
the location of the photon-sphere and hence the size of the shadow, 
demonstrating that higher-curvature corrections and the matter distribution 
leave potentially observable imprints on the propagation of light.

The dynamical response of the spacetime was examined through scalar 
perturbations using both the Pad\'{e} averaged sixth-order WKB approximation 
and time-domain evolution. The effective potential exhibits the barrier 
structure required for the WKB analysis, while the eikonal quasinormal modes 
were shown to satisfy the expected correspondence with the properties of 
unstable circular null geodesics. The WKB frequencies display good agreement 
with the time-domain results, and the agreement improves further with 
increasing multipole number, consistent with the asymptotic validity of the 
eikonal approximation. The time-domain profiles exhibit the characteristic 
quasinormal ringing followed by late-time echo-like signals associated with 
the double-barrier structure of the effective potential in the tortoise 
coordinate. Most importantly, no exponentially growing modes were observed 
during the evolution, providing strong evidence that the wormhole is 
dynamically stable against massless scalar perturbations within the parameter 
range considered.

Overall, our results demonstrate that the combined effects of the Gauss-Bonnet 
curvature corrections and the Letelier cloud of strings provide a physically 
consistent mechanism for supporting regular and traversable wormholes in 
five-dimensional spacetime. Beyond satisfying the geometrical and physical 
requirements for traversability, the solution possesses distinct optical and 
dynamical signatures that may help distinguish such wormholes from other 
compact objects. Future investigations may include gravitational and 
electromagnetic perturbations, rotating generalizations of the present 
solution and the study of their gravitational waves and observational 
signatures in higher-dimensional EGB gravity.

\section*{Acknowledgments}

UDG is thankful to the Inter-University Centre for Astronomy and Astrophysics (IUCAA), Pune, India, for awarding the Visiting Associateship of the institute.

\appendix

\end{document}